\documentclass{JFM-FLM_Au}
\usepackage{subcaption}
\usepackage{multirow}
\usepackage{xcolor}
\usepackage{overpic}
\usepackage{enumitem}
\usepackage{hyperref}
\hypersetup{
    colorlinks = true,
    linkcolor  = blue,
    filecolor  = blue,
    urlcolor   = blue,
    citecolor  = blue,
}

\lefttitle{P. Jiang \textit{et al.}}
\righttitle{Pressure fluctuations of flow over an axisymmetric hull}

\title{Experimental study on the wall-pressure fluctuations of flow over an axisymmetric hull}

\author{
Peng Jiang\aff{1},
Haoyu Zhang\aff{1},
Yi Dai\aff{1,2},
Tao Peng\aff{1,2},
Bin Xie\aff{1,2}
\and Shijun Liao\aff{1,2}
}

\affiliation{
\aff{1}School of Ocean and Civil Engineering, Shanghai Jiao Tong University, Shanghai, 200240, China
\aff{2}State Key Laboratory of Ocean Engineering, Shanghai Jiao Tong University, Shanghai, 200240, China}

\corresau{Shijun Liao, \email{sjliao@sjtu.edu.cn}}

\begin{document}
\maketitle

\begin{abstract}
Wall-pressure fluctuations on high-Reynolds-number maneuvering hulls are critical to hydroacoustics yet remain poorly understood due to a lack of validation-quality experimental data. This paper addresses this gap by presenting the first systematic, high-fidelity database of wall-pressure fluctuations on an axisymmetric hull (SUBOFF model) that covers a high Reynolds number range ($\Rey = 5.6 \times 10^6$ to $1.4 \times 10^7$) with complex maneuvering conditions (yaw and pitch). The dataset is corrected for pinhole resonance and wind tunnel background noise, and validated against canonical flat-plate data. Analysis of the straight-ahead condition reveals that increasing $\Rey$ systematically shifts spectral energy (up to 20\%) toward lower frequencies. Critically, canonical outer-variable scaling ($\sim\text{St}^{-0.7}$ for middle frequency while $\sim\text{St}^{-2.8}$ for high frequency), which holds for the fore and mid-body, breaks down in the aft-body ($\sim\text{St}^{-2}$ for middle to high frequency), revealing the dominance of non-equilibrium, pressure-gradient-driven physics. Maneuvering conditions introduce complex 3D flow phenomena: yaw maneuvers cause high-frequency suppression at the tail, consistent with shear-layer lifting by stable crossflow vortices, while the bow-down pitch maneuvers induce broadband amplification. The most critical finding occurs during the bow-up pitch maneuvers, which reveal a spatial bifurcation of the flow: the mid-body exhibits expected spectral suppression due to stable separation, while the extreme tail undergoes a massive, monotonic spectral amplification (up to 8 dB). This phenomenon, coupled with a non-monotonic transition, provides the first clear experimental evidence of a spatial flow transition from a stable separation bubble to a highly energetic region dominated by the direct impingement of 3D crossflow vortices. These findings provide foundational physical insights into non-equilibrium 3D turbulent flows and establish a crucial benchmark dataset for the validation of the flow noise prediction models.
\end{abstract}


\section{Introduction}\label{introduction}
Wall pressure fluctuations are a dominant source of flow-generated noise in wall-bounded turbulence \citep{Willmarth1975Pressure, Yang_Yang_2022}. These acoustic characteristics are important to the stealth capabilities of underwater vehicles, particularly submarines \citep{Yu2007shipNoise, Blake2017Mechanics}. Submarine noise comprises three main categories: mechanical, propeller, and hydrodynamic \citep{Yu2007shipNoise}. Hydrodynamic noise is further subdivided into flow noise and flow-induced noise. Flow noise arises primarily from turbulent boundary layer effects and flow instabilities. Flow-induced noise, in contrast, results from fluid-hull interactions, where surface pressure fluctuations excite the hull structure. Although the sound pressure level (SPL) of flow noise is generally lower than that of machinery and propeller noise, its contribution becomes significant at higher submarine speeds, often scaling with the fifth-to-sixth power of velocity. Consequently, the accurate prediction of flow noise is essential for optimizing submarine stealth performance in both civil and military applications \citep{Jiang2024OEHUll, Jiang2025PoFSUBOFF, Zhou2022suboff}.

For submerged bodies, the dominant source of the flow noise is the unsteady surface pressure fluctuations ($p'$) induced by the turbulent boundary layer. Acoustic analogies, such as the Ffowcs Williams-Hawkings (FWH) equation, provide the theoretical framework for this connection, explicitly linking the far-field radiated sound to these unsteady surface pressure fluctuations \citep{FWH1969}. The far-field acoustic signature of turbulent flows over stationary, low-speed submerged bodies is predominantly governed by the dipole term. This term, known as loading noise, directly relates the radiated acoustic pressure $\hat{p}(\boldsymbol{x}, t)$ to the unsteady surface pressure fluctuations:

\begin{equation}
4 \pi \hat{p}(\boldsymbol{x}, t) 
\approx \underbrace{\frac{1}{c_0} \frac{\partial}{\partial t} \int_{\mathbb{S}} \left[ \frac{p^{\prime} \,\widehat{n}_r}{r \,\bigl| 1 - \mathbb{M}_r \bigr|} \right]_{\mathbb{T}} dS}_{\text{Far-field term}} 
+ \underbrace{\int_{\mathbb{S}} \left[ \frac{p^{\prime} \,\widehat{n}_r}{r^2 \,\bigl| 1 - \mathbb{M}_r \bigr|} \right]_{\mathbb{T}} dS}_{\text{Near-field term}},
\label{eq:fwh_loading}
\end{equation}

where $p'$ is the fluctuating pressure on the rigid surface $S$, $\widehat{n}_r$ is the component of the surface normal unit vector in the observer's direction, $r$ is the observer distance, and $c_0$ is the speed of sound. The term $\mathbb{M}_r$ is the surface Mach number component in the $\boldsymbol{r}$ direction, and the brackets $[\cdot]_{\mathbb{T}}$ denote evaluation at the emission time. Consequently, a thorough characterization of the statistical properties, spectral content, and scaling behaviours of the wall pressure fluctuations ($p'$) is essential for developing accurate noise prediction models and effective noise reduction strategies \citep{Yang_Yang_2022}.

Given its critical role as the primary acoustic source term, extensive research has been dedicated to characterizing the properties of wall pressure fluctuations. This effort has, in turn, informed the development of predictive models. Experimental investigations into wall pressure fluctuations have systematically progressed from simple canonical flows towards more complex, application-relevant configurations. Foundational studies focused on turbulent boundary layers (TBLs) over smooth, flat plates under zero-pressure-gradient (ZPG) conditions \citep{Blake1970, Schewe1983Spatial, Goody2004}. These experiments established crucial knowledge regarding spectral shapes, which confirmed the validity of inner ($u_\tau, \nu/u_\tau$) and outer ($U_\infty, \delta$) variable scaling in different frequency regimes. Building upon the ZPG baseline, researchers introduced further complexities. The influence of pressure gradients (PGs) was investigated. Adverse pressure gradients (APGs) were found to amplify low-frequency fluctuations and affect coherence lengths. In contrast, favourable pressure gradients (FPGs) were shown to primarily impact higher frequencies. Similarly, studies on rough surfaces demonstrated significant increases in overall wall pressure fluctuation levels and alterations to spectral shapes \citep{Joseph2020JFMPlate,Joseph2017Pressure,Meyers2015Wallpressure}. Further complexity was explored through experiments on two-dimensional bluff bodies and lifting surfaces. Studies on the wall pressure fluctuations on circular cylinders have identified the association of propertoes of pressure spectrum with boundary layer separation and the periodic vortex shedding \citep{Maryami2018Cylinder, Maryami2020Turbulent}. Moreover, driven by aeroacoustic concerns (e.g. trailing edge noise), extensive experimental work on aerofoils has provided rich datasets \citep{Celik2022Experimental, Hu2021Coherence}. These datasets cover wall pressure fluctuations under various angles of attack, Reynolds numbers, Mach numbers, and inflow conditions (smooth vs. turbulent). These investigations mapped fluctuation patterns across aerofoil surfaces, highlighting intense pressures near the leading and trailing edges and in regions of strong PGs or flow separation.

The challenges associated with 2D geometries, such as flow separation and strong pressure gradients, are further intensified in three-dimensional (3D) TBLs. These flows are characteristic of complex geometries like wing-body junctions or axisymmetric bodies of revolution. Experiments in such flows have revealed the significant impact of crossflow, boundary layer skewing, and the formation of large-scale secondary structures on surface pressure statistics. For instance, \citet{ Goody2004, Goody1999Experimental} investigated wall pressure fluctuations beneath several 3D TBLs, including the separating flow on the leeward side of a 6:1 prolate spheroid at angles of attack ($\alpha = 10^\circ, 20^\circ$). Their work highlighted that scaling parameters for $p'$ spectra in these complex 3D flows, especially near crossflow separation regions, must incorporate local flow structure. This finding indicates that the `universal' 2D scaling laws may not be applicable to 3D TBLs, emphasising the significance of near-wall mean velocity gradients and Reynolds stress structures in determining high-frequency spectral levels.

A key challenge exists for axisymmetric bodies of revolution (BORs) specifically. These geometries combine transverse curvature effects with complex, spatially varying pressure gradients, especially on the forebody and stern regions. For such configurations, experimental wall pressure fluctuation data remains limited in comparison to that for simpler geometries, although early work examined curvature effects in axial flow. A significant recent contribution is the work by  \citet{Balantrapu2023APG}, who studied wall pressure fluctuations on a BOR stern under a strong, but importantly, spatially stable APG (no streamwise curvature variation) at a high Reynolds number. They found that despite the strong APG, the root-mean-square pressure scaled well with local wall shear stress ($p'_{rms} \approx 7 \tau_w$). Furthermore, the spectra collapsed effectively using wall-wake scaling ($\tau_w$ for pressure, $U_e/\delta$ for frequency), even into the viscous $f^{-5}$ region. However, the SUBOFF hull investigated herein features significant variations in both streamwise curvature and pressure gradient along its bow and stern. This presents a more complex non-equilibrium flow field where such simple scaling relationships are unlikely to hold.

Thus, a critical gap remains in the experimental characterization of wall pressure fluctuations on axisymmetric hulls, particularly under conditions relevant to real-world vehicle operation. First, high-fidelity experimental wall pressure fluctuation ($p'$) data for benchmark geometries like the DARPA SUBOFF model \citep{groves1989geometric} at high Reynolds numbers ($\Rey > 10^6$) are extremely limited, even for the baseline straight-ahead condition. Most existing SUBOFF experiments focused on mean flow or wake properties \citep{Liu1998SUBOFFexp, Jimenez2010JFMwakeflow}. Although recent high-fidelity simulations by \citet{Fan2025ScalingJFM} and \citet{Jiang2024OEHUll, Jiang2025PoFSUBOFF} have begun to explore $p'$ scaling laws under these conditions, a significant gap in experimental validation data persists. Second, and most significantly for the present work, there is a near-complete lack of systematic experimental data on $p'$ for axisymmetric hulls under maneuvering conditions (yaw and pitch) at high Reynolds numbers. Submarines rarely operate purely in straight-ahead motion. Maneuvers induce significant angles of attack, leading to complex 3D phenomena such as flow separation, asymmetric vortex shedding (e.g. crossflow vortices), and other highly unsteady turbulent structures \citep{AMIRI2019192}. These effects are expected to fundamentally alter the $p'$ field compared to the axisymmetric baseline. Understanding these effects is crucial for predicting noise signatures during operational maneuvers. The absence of experimental data can be attributed to a number of challenges to the accurate measurement of $p'$ at high speeds. These include removing the facility noise \citep{Willmarth1975Pressure}, ensuring adequate sensor spatial resolution \citep{Schewe1983Spatial}, and correcting for sensor pinhole resonance \citep{Tsuji2012Pressure}. However, this lack of validation-quality experimental data, especially for maneuvering cases, represents a major challenge for the development and verification of advanced hydroacoustic prediction models.

This study aims to address these critical gaps through a comprehensive wind tunnel experiment conducted on the SUBOFF bare hull. The primary objectives are: (i) To establish and validate the first high-fidelity experimental database of wall-pressure fluctuations on the SUBOFF hull covering a high Reynolds number range ($\Rey = 5.6 \times 10^6$ to $1.4 \times 10^7$) under both straight-ahead and complex maneuvering (yaw and pitch) conditions.
(ii) To systematically quantify the effects of Reynolds number and pressure gradients on the wall-pressure fluctuations spectra (level, shape, energy distribution) under straight-ahead conditions, with a specific focus on analyzing spectral scaling characteristics and identifying their breakdown in non-equilibrium flow regions, particularly the aft-body.
(iii) To elucidate the fundamental impact of maneuvering conditions (yaw and pitch angles) on wall-pressure fluctuations spectra, identifying the distinct physical mechanisms associated with windward versus leeward flow regimes, including 3D flow separation, crossflow vortex dynamics, and the discovery of a critical flow transition involving vortex impingement at the tail under negative pitch. By achieving these objectives, this work provides not only a unique benchmark dataset but also crucial physical insights into the complex turbulence physics governing wall-pressure fluctuations on maneuvering axisymmetric hulls at high Reynolds numbers.

The structure of the rest of the paper is outlined as follows: \S\ref{Experimental_Methodology} details the experimental methodology, including the wind tunnel facility, the SUBOFF model, the instrumentation for both mean and fluctuating pressure measurements, and signal correction methods. Subsequently, \S\ref{Results_and_Discussion} presents the main findings. It begins with a characterization of the mean flow (\S\ref{Mean_Flow}), followed by a detailed analysis of the wall-pressure spectra for the straight-ahead condition, examining the influence of Reynolds number, pressure gradients, and spectral scaling (\S\ref{Wall-Pressure_Spectra_Straight-Ahead}). The section then investigates the complex effects of maneuvering conditions, specifically yaw and pitch angles (\S\ref{Maneuvering_Conditions}). Finally, \S\ref{Concluding} summarizes the key findings of the study and discusses future work. The paper is supported by detailed appendices on data validation and uncertainty analysis.

\section{Experimental Methodology}\label{Experimental_Methodology}

\subsection{Wind tunnel facility and SUBOFF model}\label{Wind_Tunnel_SUBOFF_Model}
\begin{figure*} 
	\centering
	\begin{minipage}[c]{1\linewidth}
    	\centering
		\begin{overpic}[trim=0.3cm 0cm 0.5cm 0.3cm, clip, width=12cm]{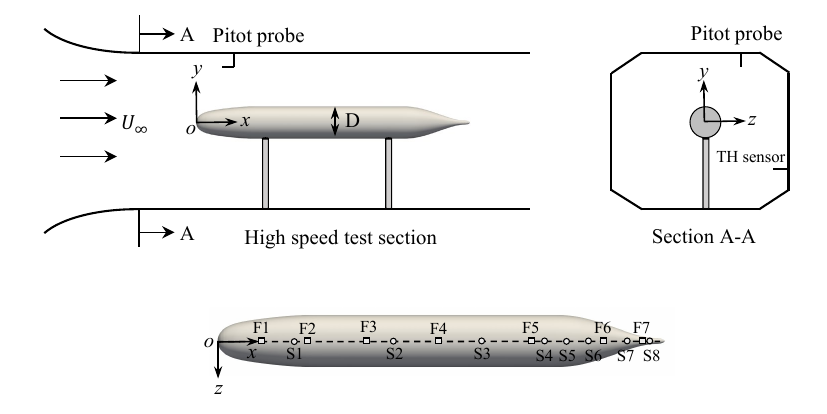}
			\put(-5,50){\color{black}{(a)}}
            \put(70,45){\color{black}{(b)}}
            \put(18,12){\color{black}{(c)}}
		\end{overpic}
	\end{minipage}
    \caption{Schematic of the experimental setup: (a) installation of the axisymmetric hull in the high-speed test section of the Shanghai Jiao Tong University wind tunnel; (b) lateral view showing Section A-A and the location of the temperature and humidity (TH) sensor; (c) arrangement of measurement points along the hull's top meridian line. Square symbols denote wall-pressure fluctuation sensors; circles denote static pressure sensors.}
    \label{fig:windtunnel_arrangement}
\end{figure*}

\begin{figure*} 
	\centering
	\begin{minipage}[c]{1\linewidth}
    	\centering
		\begin{overpic}[width=12cm]{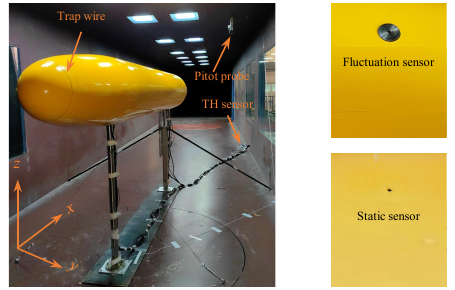}
			\put(-5,60){\color{black}{(a)}}
            \put(67,60){\color{black}{(b)}}
            \put(67,27){\color{black}{(c)}}
		\end{overpic}
	\end{minipage}
    \caption{Photographs of the experimental setup: (a) the model installed in the wind tunnel, viewed from upstream looking downstream; (b) detailed view of the wall-pressure fluctuation sensor; (c) detailed view of the static pressure sensor.}
    \label{fig:Exp_photo}
\end{figure*}

\begin{table}
\centering
\def~{\hphantom{0}}
\begin{tabular}{lcccl}
Sensor Type & Sensor ID & Axial Position  \( x \) (m) & Normalized Position \( x/L \) & Description \\ 
\multirow{8}{*}{\centering\parbox{2cm}{\centering Static \\ pressure}} 
  & S1 & 0.788 & 0.181 & Nose \\
  & S2 & 1.751 & 0.402 & Middle section \\
  & S3 & 2.618 & 0.601 & Middle section \\
  & S4 & 3.228 & 0.741 & Middle section \\
  & S5 & 3.441 & 0.790 & Aft section \\
  & S6 & 3.659 & 0.840 & Aft section \\
  & S7 & 4.038 & 0.927 & Tail \\
  & S8 & 4.260 & 0.978 & Tail end \\  \hline
\multirow{7}{*}{\centering\parbox{2cm}{\centering Fluctuation \\ pressure}} 
 & F1 & 0.457 & 0.105 & Nose \\
 & F2 & 0.904 & 0.207 & Middle section \\
 & F3 & 1.481 & 0.340 & Middle section \\
 & F4 & 2.182 & 0.501 & Middle section \\
 & F5 & 3.092 & 0.710 & Aft section \\
 & F6 & 3.795 & 0.871 & Tail \\
 & F7 & 4.164 & 0.956 & Tail end \\ 
\end{tabular}
\caption{Positions of static and surface pressure sensors along the top meridian line of the SUBOFF axisymmetric hull.}
\label{tab:sensor_positions}
\end{table}

The experimental study was conducted in the Multi-functional Wind Tunnel Facility at the State Key Laboratory of Ocean Engineering, Shanghai Jiao Tong University. This facility features three distinct test sections: a Small Test Section (High-Speed, $3 \text{ m} \times 2.5 \text{ m} \times 16 \text{ m}$, maximum wind speed $60 \text{ m/s}$), a Large Test Section (Low-Speed, $6 \text{ m} \times 3.5 \text{ m} \times 14 \text{ m}$), and an Open-Jet Test Section (Water Surface Test Section). To achieve the high Reynolds numbers required here, all tests were performed in the high-speed section.

It is important to note that in this work, wind-tunnel testing was preferred over water-tunnel testing for the following practical and theoretical reasons:
(i) Most prior experiments on wall-pressure fluctuations in turbulent boundary layers over flat plates or streamlined bodies have used wind tunnels \citep{Goody1999Experimental, Meyers2015Wallpressure, Joseph2017Pressure, Joseph2020JFMPlate, Gibeau2021JFMPlate}. This enables direct comparisons with established data and methods.
(ii) Wind tunnels simplify sensor installation (no waterproofing required), model fabrication (no leak prevention) and noise/vibration control (via acoustic isolation). Larger air sections also facilitate low blockage ratios, reducing wall interference.
(iii) Provided that the Reynolds number is matched, the resulting flow characteristics, including wall-pressure fluctuations are equivalent in air and water. Thus, wind-tunnel results can be extrapolated directly to underwater submarine flows.
(iv) Air's lower density and viscosity permit higher flow speeds for a given Reynolds number, improving resolution of high-frequency fluctuations.

The experimental model utilized in this research was a bare-hull SUBOFF axisymmetric hull \citep{groves1989geometric}, a standard benchmark for submarine hydrodynamics, at 1:24 scale. Its key parameters including overall length $L=4.356$\,m, maximum diameter $D=0.508$\,m, displacement volume $\Delta=0.699$\,m$^3$ and wetted surface area $S=5.988$\,m$^2$ are listed in table~\ref{tab:suboff_parameters}. The blockage ratio was 2.70\%, ensuring negligible wall interference.
A fully turbulent boundary layer was induced at $0.75D$ downstream of the nose using a 0.50\,mm trip wire (consistent with \citet{Jimenez2010JFMwakeflow}). The fibreglass hull (4\,mm wall thickness) had a surface tolerance of 0.2\,mm and was mounted via two 10\,cm-diameter cylindrical struts. All 15 pressure ports, eight static taps and seven dynamic transducers, were arrayed along the top side of the model. A representative setup is shown in figure~\ref{fig:Exp_photo}.

\begin{table}
\centering
\def~{\hphantom{0}}
\begin{tabular}{lcc}
\multicolumn{2}{c}{Generic submarine type}&SUBOFF bare hull\\ 
Description& Symbol& Magnitude\\
Length overall& $L$& 4.356 $\mathrm{m}$\\
Maximum hull diameter& $D$& 0.508 $\mathrm{m}$\\
Volume of displacement& $\Delta$& 0.699 $\mathrm{m^3}$\\
Wetted surface area& $S$& 5.988 $\mathrm{m^2}$\\ 
\end{tabular}
\caption{Main geometrical parameters of the axisymmetric hull of the SUBOFF model.}
\label{tab:suboff_parameters}
\end{table}

\subsection{Instrumentation and data acquisition}\label{Instrumentation_Data_Acquisition}
Static pressure measurements were conducted to characterize the mean flow field around the SUBOFF axisymmetric hull. Seven pressure taps, strategically positioned along the top meridian line of the model (as shown in figure \ref{fig:windtunnel_arrangement}(c)), were used to measure static pressure. These taps were connected to high-precision differential pressure transducers with accuracy $\pm 0.1\%$ full scale, via flexible tubing with an inner diameter of 2 mm to minimize flow disturbances. The transducers were calibrated prior to the test to ensure the measurement accuracy. The static pressure data were sampled at a frequency of 1 kHz over a duration of 90 seconds per test condition to capture steady-state flow characteristics. The data were ensemble-averaged over multiple runs to reduce random errors, and the results provide a baseline for understanding the mean flow behavior under varying Reynolds numbers and maneuvering conditions.

Surface pressure measurements were performed to characterize the unsteady wall-pressure fluctuations on the SUBOFF axisymmetric hull. To this end, seven custom-designed, high-precision wall-pressure fluctuation sensors (the CYG1506G-P4LS12C2 1/4 inch pressure-field microphones made in China), were utilized. These sensors offer a measurement range of ±2 kPa with a sensitivity of 2.5 mV/Pa. Each sensor features a measurement face diameter of 1/4 inch (6.35 mm) and includes a pinhole with 1.2 mm diameter that forms an internal pressure-sensing cavity, effectively minimizing spatial averaging effects at higher frequencies \citep{Gravante1998dPlus}. This design enabled accurate acquisition over the frequency band of interest (100\,Hz--10\,kHz). The sensors were strategically positioned along the top meridian line of the hull, as outlined in table \ref{tab:sensor_positions}, ensuring comprehensive coverage of the turbulent pressure field. To avoid flow disturbance, the transducers were flush-mounted on the hull surface. Data were acquired using an NI-395 data acquisition system, operating at a sampling frequency of 20 kHz over a 90-second recording duration per test. 

\subsection{Wall-pressure signal correction}\label{Wall-pressure_signal_correction}

Correcting the signals from the wall-pressure fluctuation sensors is a critical multi-step process. The procedure must address three primary challenges: (i) the amplitude resonance caused by the pinhole cavity, (ii) contamination from wind tunnel background noise, and (iii) the need for data smoothing to enhance spectral resolution during power spectral density computation. The procedure implemented in this investigation, which follows established methodologies \citep{Joseph2017Pressure, Joseph2020JFMPlate, Gibeau2021JFMPlate, Baars2024JFMPlate}, addresses these issues in three distinct steps, as detailed below.
\vspace{6pt}

\noindent \textit{Step 1: Microphone calibration and correction}
\vspace{6pt}

\begin{figure*} 
	\centering
	\begin{minipage}[c]{1\linewidth}
    	\centering
		\begin{overpic}[trim=0.8cm 3.5cm 0.5cm 0.7cm, clip, width=12cm]{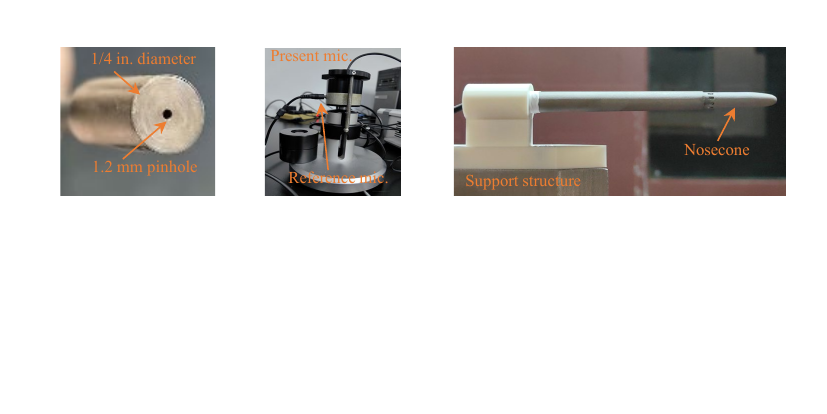}
			\put(-3,20){\color{black}{(a)}}
            \put(24,20){\color{black}{(b)}}
            \put(49,20){\color{black}{(c)}}
		\end{overpic}
	\end{minipage}
    \caption{Photographs of (a) the CYG1506G-P4LS12C2 1/4 inch pressure-field microphone with pinhole, (b) the calibration process using the HBK 9721-A acoustic sensor calibration system, and (c) the measurement of facility noise.}
    \label{fig:sensor and calibration}
\end{figure*}

As discussed in \S\ref{Instrumentation_Data_Acquisition}, each sensor features a sensing-face diameter of 6.35\,mm and a 1.2\,mm pinhole (figure~\ref{fig:sensor and calibration}(a)) to minimize spatial averaging at high frequencies \citep{Gravante1998dPlus}. However, this pinhole-cavity design forms a Helmholtz resonator, which modifies the transducer's transfer function \citep{Joseph2017Pressure, Joseph2020JFMPlate, Gibeau2021JFMPlate, Baars2024JFMPlate}. Calibration was therefore performed at the Hottinger Br{\"{u}}el \& Kj{\ae}r (HBK) calibration laboratory using the HBK 9721-A acoustic sensor calibration system (figure~\ref{fig:sensor and calibration}(b)). A frequency-domain comparison method was employed, with the HBK 4192 (a 1/4 inch pressure-field microphone with a flat frequency response from 3.15\,Hz to 20\,kHz) as the reference.

The laboratory calibration environment was controlled at a temperature of 22.4 °C, atmospheric pressure of 102.3 kPa, and humidity of 25.6\% RH. The static sensitivities of the sensors are relatively stable with a value of 2.5 mV/Pa. This calibration extended up to 20 kHz, ensuring a comprehensive assessment of the sensor response. The frequency range of interest for this study is below 10 kHz, where the effective amplification error remains within 2 dB, providing reliable measurements within this band. Based on the calibration data, the resonance peak of the pinhole cavity was identified to occur well beyond 10 kHz, thus lying outside the frequency range of interest (100 Hz--10 kHz) and minimizing its impact on the measured wall-pressure fluctuations.

To correct for dynamic response variations, particularly at high frequencies, individual transfer functions were fitted for each transducer. This yields a continuous function for consistent spectral correction across the dataset. A Fourier-transformed rational model, following \citet{Joseph2017Pressure}, was used:
\begin{equation}
C(f) = \frac{b_2 f^2 + b_1 f + b_0}{a_2 f^2 + a_1 f + a_0},
\label{eq:fitting model}
\end{equation}
where the coefficients were optimized to minimize the discrepancy between the fitted model and the observed calibration data. This continuous transfer function effectively captures the amplitude response of the pinhole resonator.

\begin{figure*} 
	\centering
	\begin{minipage}[c]{1\linewidth}
    	\centering
		\begin{overpic}[trim=0.4cm 1.5cm 0.5cm 0cm, clip, width=10cm]{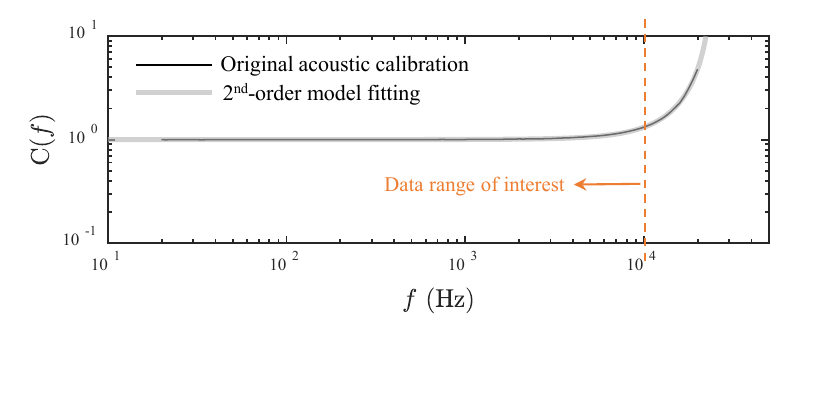}
		\end{overpic}
	\end{minipage}
    \caption{The calibrated frequency response of the pinhole resonator compared with the fitting model given by Eq.~\ref{eq:fitting model}.}
    \label{fig:frequency_response}
\end{figure*}

For illustration, the frequency response for the transducer at F4 (figure~\ref{fig:windtunnel_arrangement}(c)) is shown in figure~\ref{fig:frequency_response}. The fitted curve agrees well with the calibration data, validating the model. These fits ensure accurate correction within the 10\,kHz limit, aligning with the experimental objectives.

\vspace{6pt}
\noindent \textit{Step 2: Background noise cancellation} 
\vspace{6pt}

Accurate measurement of wall-pressure fluctuations requires suppression of wind-tunnel background noise, which can contaminate low-frequency signal components. A Wiener noise cancellation filter \citep{Hayes1996Statistical}, following \citet{Gibeau2021JFMPlate} and \citet{Baars2024JFMPlate}, was implemented. This method uses simultaneous acquisition of the wall-pressure signal and a reference noise signal, captured by a GRAS 46AE 1/2-inch free-field microphone equipped with a custom 1/2-inch nosecone to minimize turbulence-induced pressure fluctuations at the stagnation point, as illustrated in figure~\ref{fig:sensor and calibration}(c).
The Wiener filter, defined by coefficients $c$ obtained from the Wiener--Hopf equations \citep{Hayes1996Statistical}, minimizes the mean-square error between the true wall-pressure fluctuation and the filtered output. A filter order of $m=16000$ was selected iteratively to ensure spectral convergence and effective noise reduction below 10\,kHz.

\vspace{6pt}
\noindent \textit{Step 3: Data smoothing for spectral analysis}
\vspace{6pt}

To enhance spectral resolution, the corrected time-series data were segmented into 511 blocks of 8192 points each and processed with a Hanning window in MATLAB, using a 50\% overlap to improve continuity and accuracy of spectral estimates. A Fourier transform was applied to each segment, and the single-sided power spectral density was computed by averaging the transforms. This approach ensures a robust representation of the frequency content, enabling clear identification of turbulence-induced pressure features within the 10\,kHz range.

\subsection{Validation of wall-pressure signal correction}\label{correction_methods_Validation}

To validate the sensor calibration and noise cancellation methods described in \S\ref{Wall-pressure_signal_correction}, a smooth flat-plate turbulent boundary layer experiment was conducted in the same wind tunnel at a freestream velocity of $U_\infty = 30$\,m/s. The setup comprised a 3.25\,m long, 0.57\,m wide ABS plastic plate with a 1\% thickness taper at the leading edge to induce transition, mounted horizontally at zero incidence. Wall-pressure measurements were obtained using a flush-mounted CYG1506G-P4LS12C2 microphone at $x = 1.80$\,m from the leading edge, where the boundary layer was fully developed. The processed spectra were compared with benchmark data from \citet{Joseph2020JFMPlate}, who used similar pinhole microphones in an anechoic wind tunnel under identical freestream conditions ($U_\infty = 30$\,m/s).

\begin{figure*} 
\centering
	\begin{minipage}[c]{1\linewidth}
    \centering
	   \begin{overpic}[width=6cm]{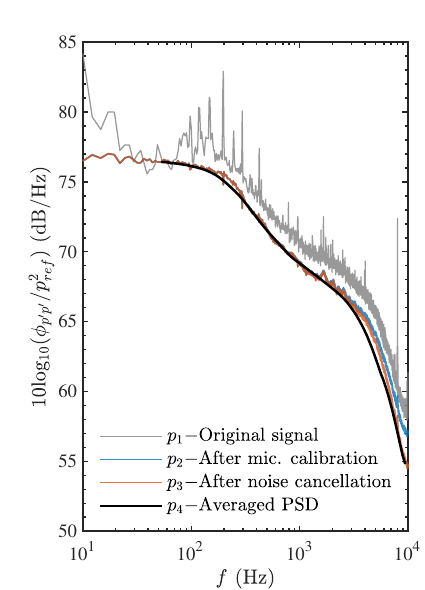}
			\put(3,94){\color{black}{(a)}}
	   \end{overpic}
	   \begin{overpic}[width=6cm]{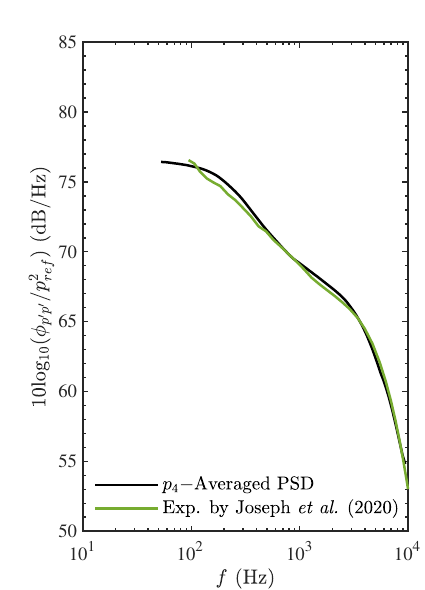}
			\put(3,94){\color{black}{(b)}}
	   \end{overpic}
	\end{minipage}
    \caption{(a) Step-by-step processing of the wall-pressure signal from the flat-plate TBL experiment at $U_\infty = 30$ m/s: $p_1$, original signal; $p_2$, after noise cancellation; $p_3$, after microphone calibration; $p_4$, averaged PSD. (b) Comparison of the averaged PSD ($p_4$) with data from \citet{Joseph2020JFMPlate}.}
    \label{fig:validation_psd}
\end{figure*}

Figure~\ref{fig:validation_psd}(a) illustrates the correction process for a representative signal: the original signal ($p_1$) exhibits facility noise contamination, which is handled by noise cancellation ($p_2$), refined by microphone calibration ($p_3$), and finalized as the averaged PSD ($p_4$). Figure~\ref{fig:validation_psd}(b) shows that $p_4$ closely matches Joseph \textit{et al.}'s smooth-wall spectrum \citep{Joseph2020JFMPlate}, with deviations below 3\,dB across the low- to mid-frequency range ($<10$\,kHz), despite the conventional tunnel's higher ambient noise. This agreement, consistent with the expected slopes of the two-layer model, confirms that the correction methods effectively eliminate facility-specific noise and the effect of the cavity resonance. Thus, the SUBOFF hull results in the main text accurately represent turbulence-induced pressure fluctuations, free from environmental or instrumental biases.

\subsection{Summary of data set}\label{Summary_data_set}
The experiments were structured around two primary parameter sets: Reynolds number and maneuvering conditions. This test matrix was designed to provide comprehensive coverage of both baseline turbulent flow and dynamic maneuvering effects. These conditions cover typical submarine maneuvers, capturing the impact of complex flow phenomena such as separation and reattachment on the wall-pressure spectra.

\begin{enumerate}[label=(\roman*)]
    \item Reynolds Number: Measurements were conducted at four high-$\Rey$ conditions based on model length $L$: $\Rey = 5.6 \times 10^6$, $8.4 \times 10^6$, $1.2 \times 10^7$, and $1.4 \times 10^7$. These correspond to freestream velocities $U_\infty$ of 20, 30, 43, and 50 m/s, respectively.
    
    \item Maneuvering Conditions: The effects of hull orientation were evaluated across a matrix of yaw and pitch angles, in addition to the baseline straight-ahead case:
    \begin{enumerate}[label=(\alph*)]
        \item Straight-ahead: $0^\circ$ (Baseline condition).
        \item Yaw Angles: $3^\circ$ and $6^\circ$. Symmetry was assumed for horizontal maneuvers due to flow invariance.
        \item Pitch Angles: $\pm 3^\circ$ and $\pm 6^\circ$, representing both bow-down (positive) and bow-up (negative) motions.
    \end{enumerate}
\end{enumerate}

\section{Results and discussion}\label{Results_and_Discussion}
\subsection{Mean flow characteristics}\label{Mean_Flow}

The mean flow field over the axisymmetric hull was measured for Reynolds numbers (based on model length) ranging from $5.6 \times 10^6$ to $1.4 \times 10^7$. Static pressure taps along the dorsal meridian (table~\ref{tab:sensor_positions}) provided the data, expressed as the pressure coefficient:
\begin{equation}
    C_p = \frac{p - p_\infty}{\frac{1}{2} \rho U_\infty^2},
\end{equation}
where $p$ is the local static pressure, $p_\infty$ the freestream static pressure, $\rho$ the air density, and $U_\infty$ the freestream velocity. The highest Reynolds number exceeds prior laboratory studies of similar geometries by an order of magnitude, enabling detailed evaluation of high-$\Rey$ effects.

\begin{figure*} 
	\centering
	\begin{overpic}[trim=3.5cm 0cm 3cm 0cm, clip, width=9cm]{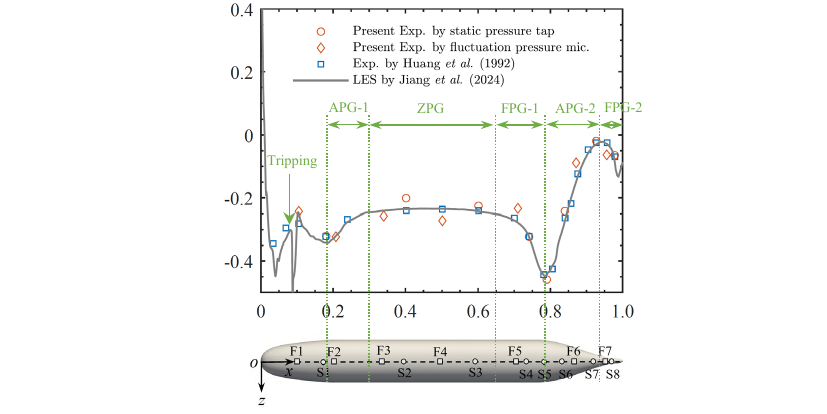}
		\put(-2,55){{\color{black}{\rotatebox{90}{$C_p$}}}}
		\put(50,20){{\color{black}{$x/L$}}}
	\end{overpic}
	\caption{Mean pressure coefficient ($C_p$) distribution along the SUBOFF hull at $\Rey = 1.2 \times 10^7$ (straight-ahead condition) compared with experimental data from \citet{Liu1998SUBOFFexp} and LES from \citet{Jiang2024OEHUll}, with schematic of flow region division.}
	\label{fig:mean-pressure-comparison-and-flow-region-division}
\end{figure*}

\begin{figure*} 
	\centering
	\begin{overpic}[width=9cm]{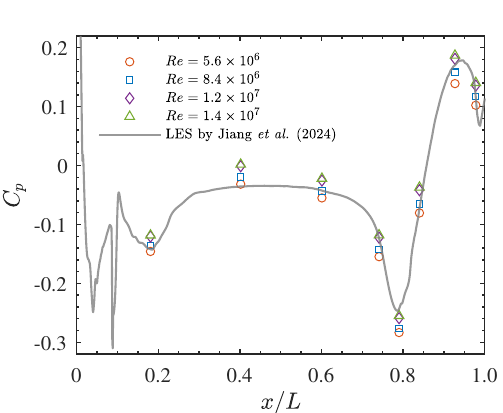}
	\end{overpic}
	\caption{Mean pressure coefficient ($C_p$) along the SUBOFF hull under straight-ahead conditions for $\Rey = 5.6 \times 10^6$ to $1.4 \times 10^7$, compared with LES from \citet{Jiang2024OEHUll}.}
	\label{fig:mean_cp_straight_ahead}
\end{figure*}

\begin{figure*} 
	\centering
	\begin{minipage}{0.48\linewidth}
		\centering
		\begin{overpic}[width=6cm]{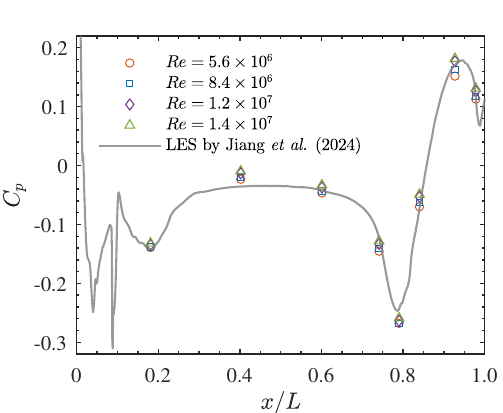}
			\put(0,75){\color{black}{(a)}}
		\end{overpic}
	\end{minipage}
	\begin{minipage}{0.48\linewidth}
		\centering
		\begin{overpic}[width=6cm]{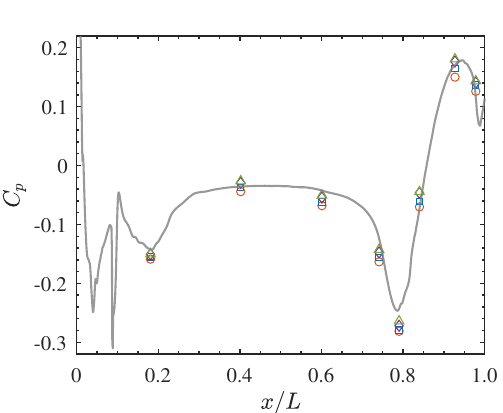}
			\put(0,75){\color{black}{(b)}}
		\end{overpic}
	\end{minipage}
	\vspace{0.5cm}
	\begin{minipage}{0.48\linewidth}
		\centering
		\begin{overpic}[width=6cm]{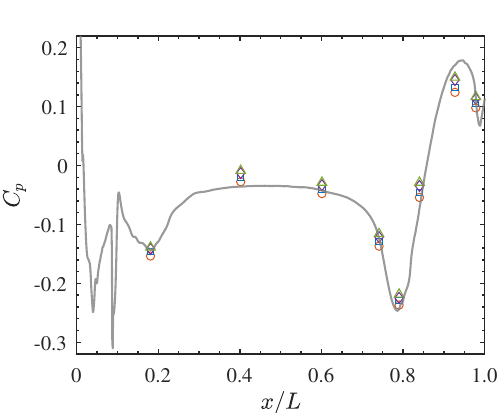}
			\put(0,75){\color{black}{(c)}}
		\end{overpic}
	\end{minipage}
	\begin{minipage}{0.48\linewidth}
		\centering
		\begin{overpic}[width=6cm]{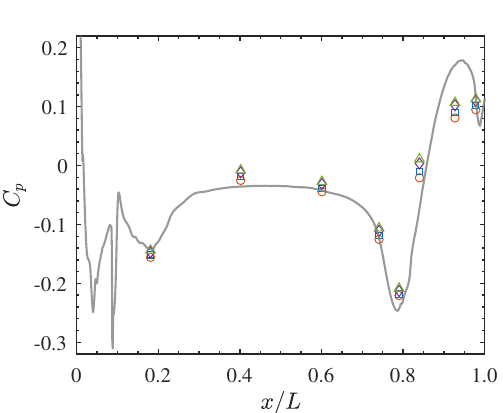}
			\put(0,75){\color{black}{(d)}}
		\end{overpic}
	\end{minipage}
	\vspace{0.5cm}
	\begin{minipage}{0.48\linewidth}
		\centering
		\begin{overpic}[width=6cm]{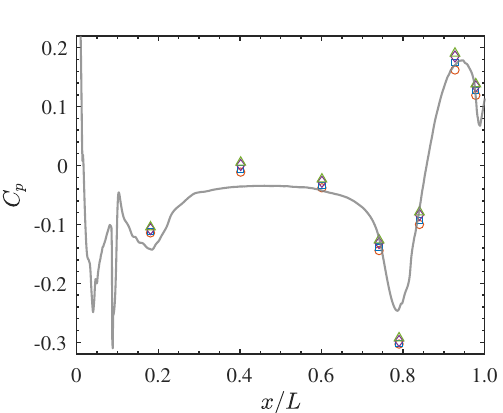}
			\put(0,75){\color{black}{(e)}}
		\end{overpic}
	\end{minipage}
	\begin{minipage}{0.48\linewidth}
		\centering
		\begin{overpic}[width=6cm]{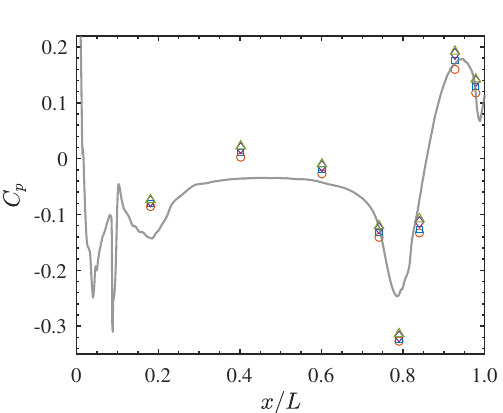}
			\put(0,75){\color{black}{(f)}}
		\end{overpic}
	\end{minipage}
	\caption{Mean pressure coefficient ($C_p$) along the SUBOFF hull at various yaw and pitch angles across $\Rey = 5.6 \times 10^6$ to $1.4 \times 10^7$: (a) yaw $3^\circ$, (b) yaw $6^\circ$, (c) pitch $-3^\circ$, (d) pitch $-6^\circ$, (e) pitch $+3^\circ$, (f) pitch $+6^\circ$. All panels include LES from \citet{Jiang2024OEHUll} at $\Rey = 1.2 \times 10^7$ (straight-ahead) for reference.}
	\label{fig:mean_cp_maneuvering_angles}
\end{figure*}

Figure~\ref{fig:mean-pressure-comparison-and-flow-region-division} shows the $C_p$ distribution along the dorsal meridian at $\Rey = 1.2 \times 10^7$ (straight-ahead condition). Near the nose, $C_p$ is elevated due to stagnation, followed by a sharp decrease as the flow accelerates over the forebody ($x/L<0.2$). A short adverse pressure gradient region (APG-1, $\mathrm{d}p/\mathrm{d}x > 0$) appears just upstream of the parallel section ($0.2 < x/L < 0.3$). The pressure then remains nearly constant across the parallel middle body ($0.3 < x/L < 0.65$, ZPG), decreases further at the tail onset ($0.65 < x/L < 0.8$, FPG-1, $\mathrm{d}p/\mathrm{d}x < 0$), rises gradually in the stern taper ($0.8 < x/L < 0.95$, APG-2), and ends with a short favorable gradient near the trailing edge ($x/L > 0.95$, FPG-2). This profile reflects the hull geometry: the streamlined nose induces stagnation and rapid acceleration, the parallel middle body sustains near-zero pressure gradient, and the stern taper drives initial acceleration (convex curvature) followed by deceleration and pressure recovery (concave curvature). The latter region is known for rapid boundary-layer thickening due to strong cross-layer pressure variations \citep{Patel1974SUBOFFExp, Liu1998SUBOFFexp}. The present data align closely with measurements by \citet{Liu1998SUBOFFexp} and high-fidelity LES by \citet{Jiang2024OEHUll} at the same $\Rey$, confirming experimental fidelity. {\color{black}Moreover, as shown in figure~\ref{fig:mean_cp_straight_ahead},} under straight-ahead conditions, the $C_p$ distribution remains qualitatively consistent across all tested Reynolds numbers (figure~\ref{fig:mean_cp_straight_ahead}), indicating a fully developed turbulent boundary layer. {\color{black}Nevertheless,} as $\Rey$ increases from $5.6 \times 10^6$ to $1.4 \times 10^7$, the curves shift subtly upward, with the minimum $C_p$ moving forward by $\Delta(x/L) \approx 0.02$ and the tail adverse gradient weakening by $\sim$10\%. This trend, also observed in prior axisymmetric studies \citep{Liu1998SUBOFFexp}, arises from the reduced relative thickness of the boundary layer at higher $\Rey$. The diminished displacement effect allows the external flow to more closely follow the geometric contour, reducing acceleration over the hull and yielding higher local pressures (less negative $C_p$) via Bernoulli's principle. {\color{black}As a result,} the flow approaches the inviscid slender-body solution, with improved pressure recovery in the stern-consistent with high-$\Rey$ axisymmetric wake behavior, where low-$\Rey$ boundary layers are more prone to adverse-gradient separation \citep{Jimenez2010JFMwakeflow}.

{\color{black}Turning to maneuvering conditions,} under maneuvering conditions, the mean pressure distributions deviate from the symmetric straight-ahead case, as shown in figure~\ref{fig:mean_cp_maneuvering_angles}. For yaw maneuvers, $C_p$ profiles vary across hull regions, with Reynolds number effects differing from the straight-ahead configuration. In the nose ($x/L < 0.1$) and parallel middle body ($0.1 < x/L < 0.7$), $C_p$ changes minimally as $\Rey$ increases from $5.6 \times 10^6$ to $1.4 \times 10^7$, owing to yaw-induced crossflow governed by inviscid outer flow over the attached boundary layer, rendering it largely $\Rey$-insensitive. {\color{black}In contrast,} conversely, in the tail adverse pressure gradient region ($x/L > 0.7$), $C_p$ rises by up to 10\% with increasing $\Rey$, reflecting enhanced separation resistance and improved pressure recovery due to a thinner boundary layer in the concave taper. {\color{black}A similar but distinct pattern emerges in} for pitch maneuvers, positive pitch denotes nose-down (increased angle of attack on the lower surface), while negative pitch indicates nose-up. Reynolds number variations exert a weaker influence on $C_p$ than in straight-ahead conditions, as vertical asymmetry dominates viscous effects. In negative pitch cases ($-3^\circ$ and $-6^\circ$), the tail adverse gradient flattens, with larger angles reducing it by up to 10\% by alleviating upper-surface deceleration in the concave tail and mitigating separation via favorable curvature. {\color{black}Conversely, in} for positive pitch ($+3^\circ$ and $+6^\circ$), $C_p$ decreases by $\sim$5\% in the fore- and mid-body due to accelerated upper-surface flow from effective camber, yielding thinner boundary layers and stronger suction; however, the tail gradient steepens, promoting separation.

\subsection{Wall-Pressure spectra: straight-ahead condition}\label{Wall-Pressure_Spectra_Straight-Ahead}

The wall-pressure spectrum is important for analyzing turbulent fluctuations that generate flow noise of underwater vehicles. This section tends to analyse the wall-pressure spectra for the straight-ahead condition. As shown in figure~\ref{fig:mean-pressure-comparison-and-flow-region-division}, the seven fluctuating pressure sensors (F1--F7) are strategically positioned to capture the distinct flow regimes identified from the mean $C_p$ distribution. Sensor F1, adjacent to the tripping wire, monitors the transitional and developing boundary layer. Sensor F2 is located in the initial adverse pressure gradient (APG-1) region. Sensors F3 and F4 are situated within the parallel middle body's zero-pressure-gradient (ZPG) region, with F4 ($x/L = 0.501$) serving as the primary reference for an equilibrium boundary layer. Further aft, sensor F5 is in the favourable pressure gradient (FPG-1) region at the tail onset. Finally, sensors F6 (in APG-2) and F7 (in FPG-2) are positioned to capture the complex, non-equilibrium flow dynamics in the stern, including strong adverse gradients and pressure recovery. This arrangement facilitates a systematic analysis of how pressure gradients and flow history influence the wall-pressure fluctuations.

\subsubsection{Effect of the Reynolds number}\label{Reynolds_number_effect}

\begin{figure*} 
\centering
	\begin{minipage}[c]{1\linewidth}
    \centering
	   \begin{overpic}[width=5.5cm]{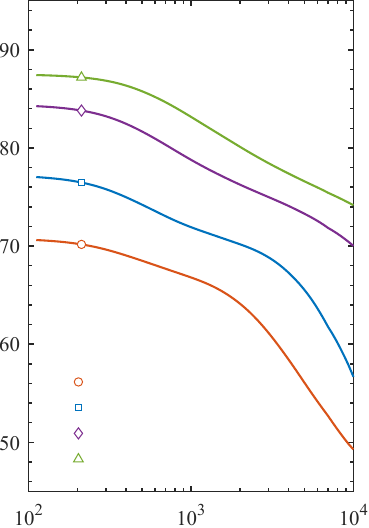}
			\put(-5,98){\color{black}{(a)}}
            \put(-7,30){{\color{black}{\rotatebox{90}{$10 \log_{10}(\phi_{p^\prime p^\prime}/p^2_\text{ref})\, \text{(dB/Hz)}$}}}}
            \put(30,-5){{\color{black}{$f$ (Hz)}}}
            \put(10,94){\color{black}\small{F3}}
            \put(18,26){\color{black}\footnotesize{$\Rey = 5.6 \times 10^6$}}
            \put(18,21){\color{black}\footnotesize{$\Rey = 8.4 \times 10^6$}}
            \put(18,16){\color{black}\footnotesize{$\Rey = 1.2 \times 10^7$}}
            \put(18,11.5){\color{black}\footnotesize{$\Rey = 1.4 \times 10^7$}}
            \put(45,94){\color{black}\footnotesize{dB re 20 $\mu$Pa}}
	   \end{overpic}
       \hspace{0.1cm}
	   \begin{overpic}[width=5.5cm]{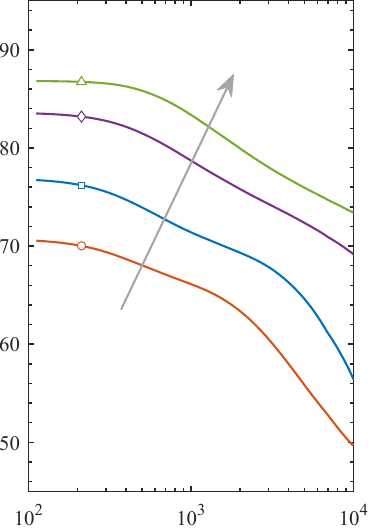}
			\put(-2,98){\color{black}{(b)}}
            \put(10,94){\color{black}\small{F4}}
            \put(30,-5){{\color{black}{$f$ (Hz)}}}
            \put(40,90){\color{black}\footnotesize{$\Rey$ increase}}
	   \end{overpic}
	\end{minipage}
    \vspace{0.5cm}
    \caption{Comparison of power spectral density of wall-pressure fluctuations at sensors (a) F3 and (b) F4 under straight-ahead conditions for Reynolds numbers ranging from $5.6 \times 10^6$ to $1.4 \times 10^7$. Note that both sensors are located in the ZPG region. The PSDs are illustrated in dB/Hz relative to a reference pressure of 20 $\mu$Pa.}
    \label{fig:F3-F4-0deg-Re}
\end{figure*}
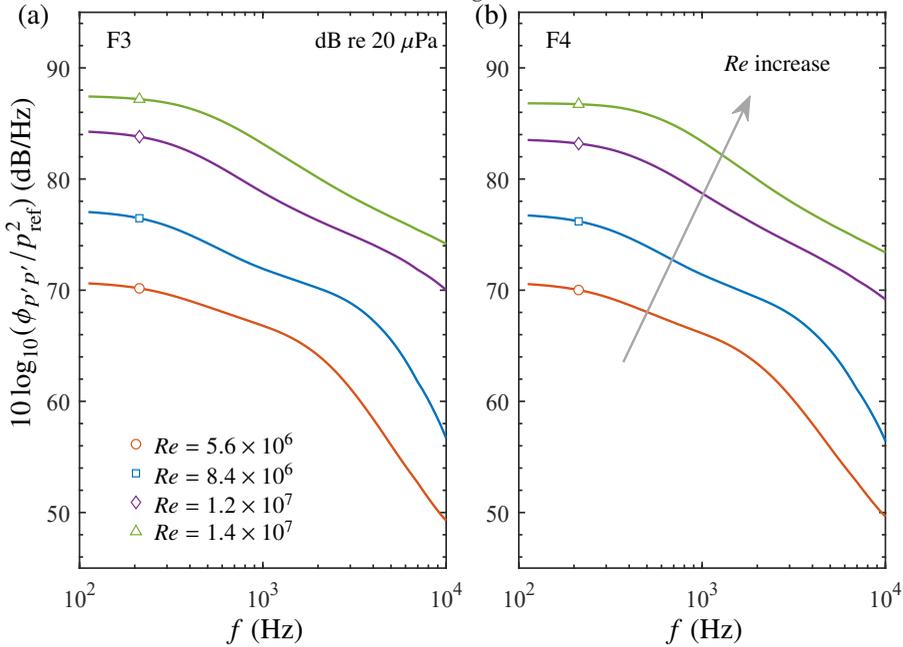

Figure~\ref{fig:F3-F4-0deg-Re} illustrates the effect of the Reynolds number on the wall-pressure PSD in the ZPG region, using data from sensors F3 and F4. Both sensors, located in the parallel middle body, show nearly identical spectral characteristics at any given $\Rey$, confirming the homogeneity of the equilibrium boundary layer in this section. A systematic $\Rey$ effect is evident. As the Reynolds number increases from $\Rey = 5.6 \times 10^6$ to $1.4 \times 10^7$, the PSD levels across the low-to-mid-frequency range (below $\approx 1$ kHz) exhibit a slight but consistent upward shift. This indicates a modest amplification of pressure-fluctuation energy with increasing $\Rey$. This phenomenon is likely due to the thinning of the boundary layer at higher Reynolds numbers. A thinner boundary layer can increase the relative influence of large-scale, energy-containing structures in the outer layer (e.g. very-large-scale motions), which are known to contribute significantly to low-frequency wall pressure. While the absolute spectral levels shift, the overall spectral shape-including the characteristic roll-off slopes-remains consistent across the tested Reynolds number range. This self-similarity in shape suggests that the fundamental structure of the equilibrium turbulent boundary layer is preserved, even as the energetic contribution from outer-layer structures scales with $\Rey$.

\subsubsection{Influence of the boundary-layer trip}\label{boundary_tripping_influence}
\begin{figure*} 
\centering
	\begin{minipage}[c]{1\linewidth}
    \centering
	   \begin{overpic}[width=5.5cm]{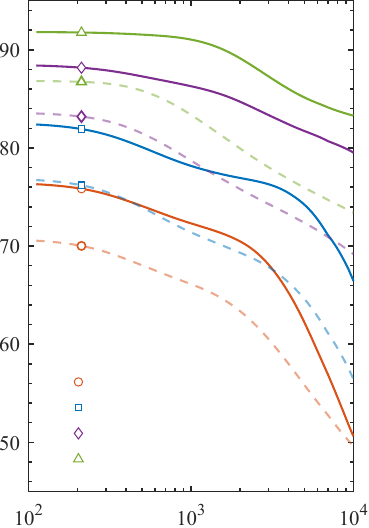}
            \put(-7,30){{\color{black}{\rotatebox{90}{$10 \log_{10}(\phi_{p^\prime p^\prime}/p^2_\text{ref})\, \text{(dB/Hz)}$}}}}
            \put(30,-5){{\color{black}{$f$ (Hz)}}}
            \put(10,95){\color{black}\small{F1}}
            \put(18,26){\color{black}\footnotesize{$\Rey = 5.6 \times 10^6$}}
            \put(18,21){\color{black}\footnotesize{$\Rey = 8.4 \times 10^6$}}
            \put(18,16){\color{black}\footnotesize{$\Rey = 1.2 \times 10^7$}}
            \put(18,11.5){\color{black}\footnotesize{$\Rey = 1.4 \times 10^7$}}
            \put(45,94){\color{black}\footnotesize{dB re 20 $\mu$Pa}}
	   \end{overpic}
	\end{minipage}
    \vspace{0.5cm}
    \caption{Comparison of the PSD of wall-pressure fluctuations at sensor F1 (behind the tripping wire, solid lines) and F4 (in the zero-pressure-gradient region, dashed lines) under straight-ahead conditions for Reynolds numbers from $5.6 \times 10^6$ to $1.4 \times 10^7$.}
    \label{fig:F1-F4-0deg-Re}
\end{figure*}
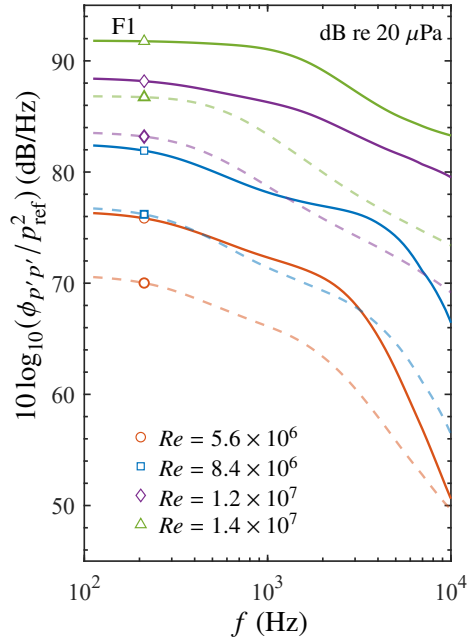

Figure~\ref{fig:F1-F4-0deg-Re} compares the power spectral density of wall-pressure fluctuations at sensors F1 (behind the tripping wire) and F4 (in the zero-pressure-gradient region) to investigate the effect of tripping wire on wall-pressure spectra. The results reveal a significant difference in wall-pressure fluctuations at F1 compared to F4. In detail, the pressure spectrum at F1 is characterized by: 
(i) Higher pressure fluctuations: The PSD at F1 is significantly higher than at F4, especially in the low- and mid-frequency ranges, indicating the tripping wire introduces disturbances that amplify pressure fluctuations at the wall.
(ii) Different spectral slopes: At F1, the spectral slope is steeper than at F4, especially in the higher frequency range. While the PSD at F4 decays gradually across frequencies, the PSD at F1 exhibits a more rapid decay, highlighting the influence of the tripping wire not only on low-frequency fluctuations but also on the frequency characteristics of the wall-pressure spectrum.
(iii) Shifted frequency peaks: The energy-containing frequency ranges at F1 are shifted towards lower frequencies compared to F4, which is consistent with the large-scale flow structures created by the tripping wire, leading to the dominance of lower-frequency turbulent structures in the flow.
These observations indicate that the trip introduces substantial, non-equilibrium disturbances that locally amplify pressure fluctuations and fundamentally alter the spectral energy distribution. Nevertheless, as established in \S\ref{Reynolds_number_effect}, the near-identical spectra at sensors F3 and F4 confirm that the flow rapidly evolves into a canonical, fully-development turbulent boundary layer. This demonstrates that the boundary-layer trip is effective in achieving transition, while its severe disruptive effects remain localized to its immediate vicinity. This finding is consistent with the numerical simulations of \citet{MorseMahesh2023Trippingeffects}, who also reported the significant impact of tripping on the flow development over the SUBOFF hull.

\subsubsection{Influence of pressure gradients}\label{pressure_gradients_influence}
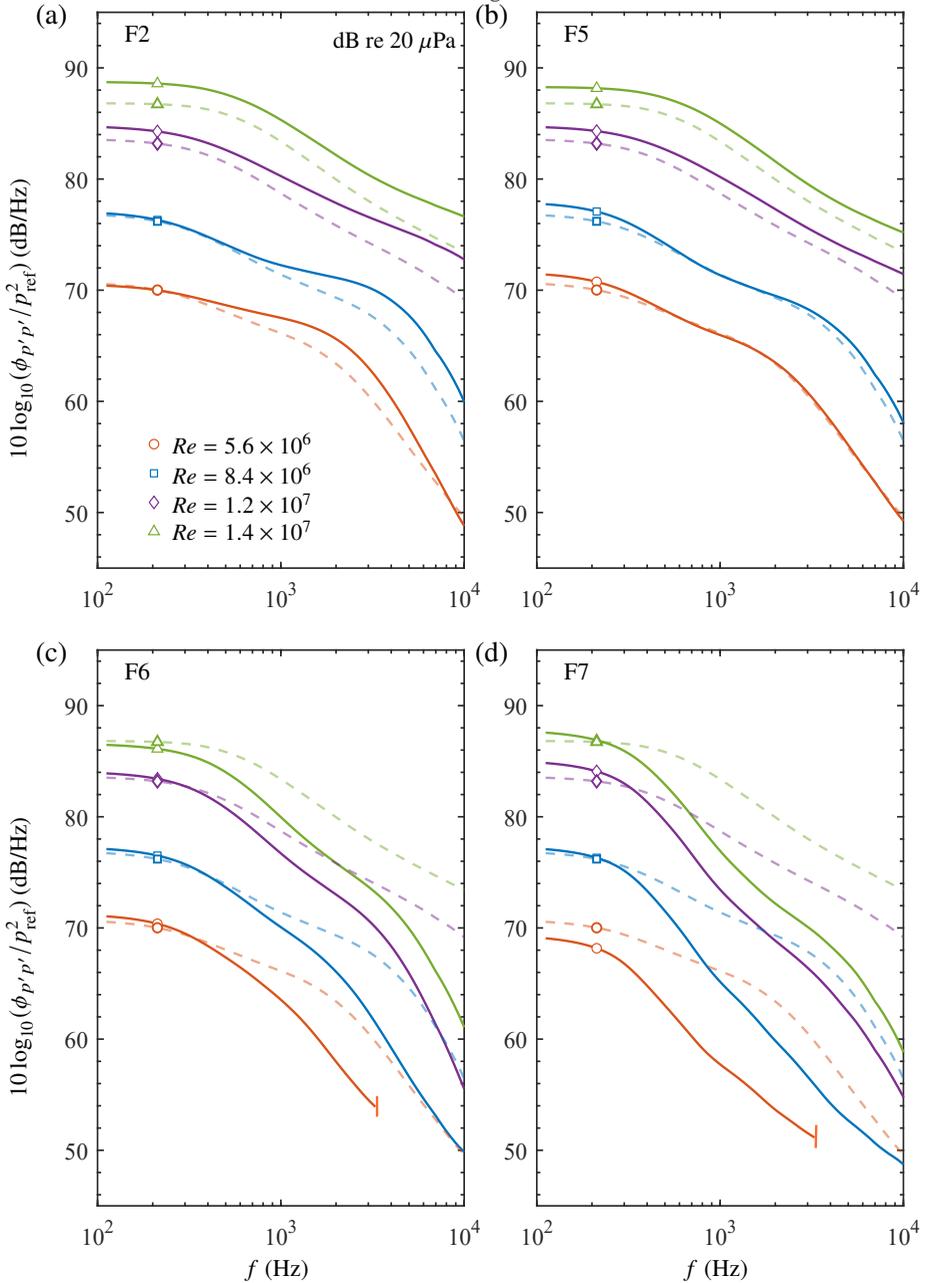
\begin{figure*} 
    \centering
	\begin{minipage}[c]{1.0\linewidth}
    	\centering
    \begin{overpic}[width=5.5cm]{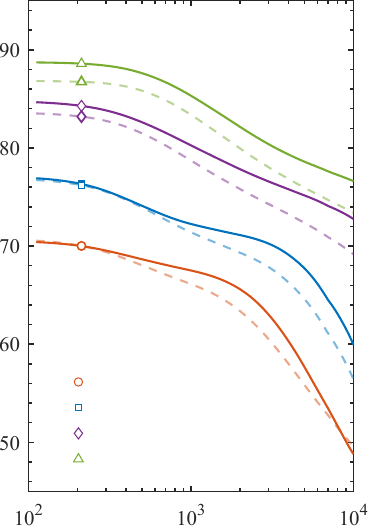}
        \put(-5,98){\color{black}{(a)}}
        \put(10,95){\color{black}\small{F2}}
        \put(45,94){\color{black}\footnotesize{dB re 20 $\mu$Pa}}
        \put(18,26){\color{black}\footnotesize{$\Rey = 5.6 \times 10^6$}}
        \put(18,21){\color{black}\footnotesize{$\Rey = 8.4 \times 10^6$}}
        \put(18,16){\color{black}\footnotesize{$\Rey = 1.2 \times 10^7$}}
        \put(18,11.5){\color{black}\footnotesize{$\Rey = 1.4 \times 10^7$}}
        \put(-10,25){\small{\color{black}{\rotatebox{90}{$10 \log_{10}(\phi_{p^\prime p^\prime}/p^2_\text{ref})\, \text{(dB/Hz)}$}}}}
    \end{overpic}
    \hspace{0.1cm}
    \begin{overpic}[width=5.5cm]{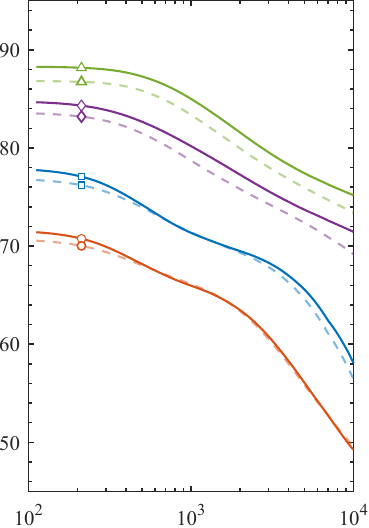}
        \put(-5,98){\color{black}{(b)}}
        \put(10,95){\color{black}\small{F5}}
    \end{overpic}
    \end{minipage}

    \vspace{0.5cm}
    
	\begin{minipage}[c]{1.0\linewidth}
    	\centering
    \begin{overpic}[width=5.5cm]{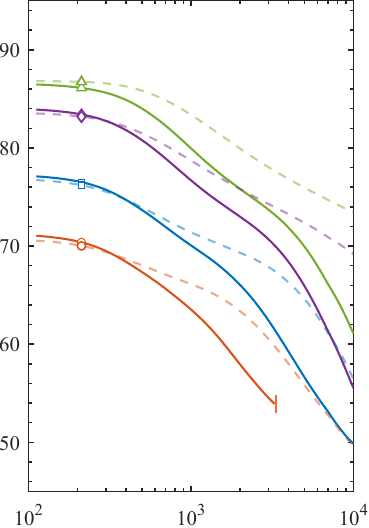}
        \put(-5,98){\color{black}{(c)}}
        \put(-10,25){\small{\color{black}{\rotatebox{90}{$10 \log_{10}(\phi_{p^\prime p^\prime}/p^2_\text{ref})\, \text{(dB/Hz)}$}}}}
        \put(30,-5){\small{\color{black}{$f$ (Hz)}}}
        \put(10,95){\color{black}\small{F6}}
    \end{overpic}
    \hspace{0.1cm}
    \begin{overpic}[width=5.5cm]{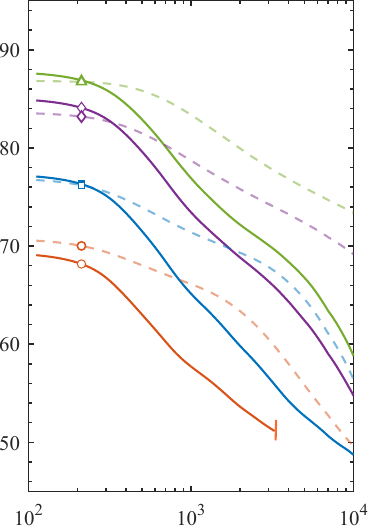}
        \put(-5,98){\color{black}{(d)}}
        \put(30,-5){\small{\color{black}{$f$ (Hz)}}}
        \put(10,95){\color{black}\small{F7}}
    \end{overpic}
    \end{minipage}
    \vspace{0.5cm}
    \caption{Comparison of the PSD of wall-pressure fluctuations at four sensors located in pressure-gradient regions-(a) F2 (APG-1), (b) F5 (FPG-1), (c) F6 (APG-2), and (d) F7 (FPG-2)--with F4 in the ZPG region used as the reference. Solid lines denote the local sensor, and the dashed line denotes F4. Results are for straight-ahead conditions over $\Rey=5.6\times10^{6}$ to $1.4\times10^{7}$, in dB/Hz re $20\,\mu$Pa.}
    \label{fig:F2_F5_F6_F7_0deg_Re_pressure_gradients}
\end{figure*}

Figure~\ref{fig:F2_F5_F6_F7_0deg_Re_pressure_gradients} compares the power spectral density of wall-pressure fluctuations at sensors F2, F5, F6, and F7 with the zero-pressure-gradient (ZPG) reference at F4. The comparison is under straight-ahead conditions across $\Rey = 5.6 \times 10^6$ to $1.4 \times 10^7$. Solid lines represent the local sensor; dashed lines, F4 at the same $\Rey$. Absolute PSD levels rise with increasing $\Rey$ (curves shift upward from orange to green). This reflects the increase in dynamic pressure. However, relative differences from the ZPG reference reveal complex, $\Rey$-dependent pressure-gradient effects.

\begin{enumerate}[label=(\roman*)]
\item Sensor F2 (APG-1, $x/L = 0.207$): This sensor, located in the initial adverse pressure gradient region, shows a highly complex relationship with the reference PSD F4. 
At lower Reynolds numbers ($\Rey \le 8.4 \times 10^6$), the PSD of F2 (solid lines) is slightly higher than F4's PSD (dashed lines) across the entire frequency range. Crucially, the curves show that the low and mid-frequency levels of F2 and F4 are relatively close, while the difference becomes more apparent at high frequencies. This suggests that at moderate $\Rey$, the initial APG mainly excites the smaller-scale turbulent structures near the wall.
At higher Reynolds numbers ($\Rey \ge 1.2 \times 10^7$), the PSD of F2 becomes significantly higher than F4 across the entire frequency spectrum. This dramatic enhancement across all frequencies at high $\Rey$ indicates that the APG effect amplifies both the large-scale structures (low frequency) and the intense small-scale structures (high frequency) in the thinner boundary layer inherent to higher $\Rey$ flow, a mechanism strongly related to flow curvature and enhanced shear production.

\item Sensor F5 (FPG-1, $x/L = 0.710$): Located in the favorable pressure gradient region at the parallel middle body's end.
At low and moderate Reynolds numbers ($\Rey \le 8.4 \times 10^6$), the PSD curves of F5 and F4 are almost entirely coincident, indicating that the FPG effect exactly counterbalances any local deviations from ZPG turbulence.
At high Reynolds numbers ($\Rey \ge 1.2 \times 10^7$), the PSD of F5 shows a clear overall upward shift relative to F4, with the PSD level increasing by approximately 1 dB to 2 dB.

\item Sensor F6 (APG-2, $x/L = 0.871$) and F7 (FPG-2, $x/L = 0.956$): These two sensors, located in the critical stern region (near tail), exhibit a shared characteristic: their PSD levels are uniformly lower than the ZPG reference F4, particularly at mid-to-high frequencies. The low-frequency plateau (initial magnitude at $f \approx 100 \text{ Hz}$) for both F6 and F7 remains essentially consistent with the F4 reference, indicating that the large-scale energy contribution is similar to the ZPG region at the same $\Rey$.
However, for mid-to-high frequencies, the PSD levels of F6 and F7 are clearly suppressed compared to F4, with F7 showing the most pronounced high-frequency attenuation. This suppression, especially strong at F7 (FPG-2), is attributed to the combined effects of hull closure: the overall flow around the stern is highly unsteady, the wake at both F6 and F7 lead to boundary-layer damping of the small-scale turbulence, reducing the high-frequency wall-pressure component. The greater suppression at F7, located further downstream and influenced by stronger FPG before the wake, confirms this stabilizing tendency on small-scale motions.
\end{enumerate}
This detailed analysis confirms that PSD levels universally increase with $\Rey$. However, the relative spectral shape changes markedly depending on the pressure gradient regime. The forward APG (F2) shows high-$\Rey$ amplification across all frequencies, while the stern APG (F6) and FPG (F7) display a unique suppression of mid-to-high frequency content, suggesting that the local non-equilibrium effects and the onset of wake dynamics strongly modulate the small-scale wall turbulence in the afterbody region.

\subsubsection{Spectral scaling characteristics}\label{Scaling_characteristics}
\begin{figure*} 
    \centering
	\begin{minipage}[c]{1.0\linewidth}
    	\centering
    \begin{overpic}[width=4cm]{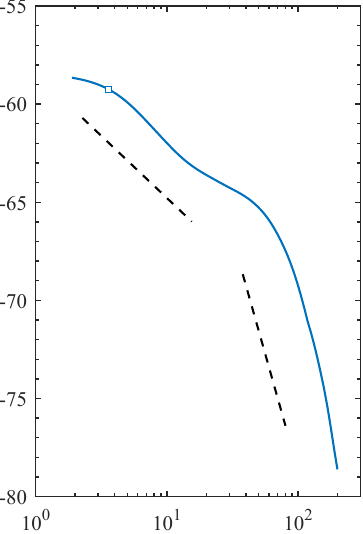}
        \put(-8,98){\color{black}{(a)}}
        \put(10,93){\color{black}\small{F2 (in APG-1)}}
        \put(10,15){\color{black}\footnotesize{$\Rey = 8.4 \times 10^6$}}
        \put(10,10){\color{black}\footnotesize{$U_\infty = 30\, \text{m/s}$}}
        \put(-10,20){{\color{black}\footnotesize{\rotatebox{90}{$10\log_{10}((\phi_{p^\prime p^\prime}/(\rho U_\infty^2)^2)(U_\infty/D))$}}}}
        \put(15,57){\color{black}\footnotesize{$\sim\text{St}^{-0.7}$}}
        \put(32,30){\color{black}\footnotesize{$\sim\text{St}^{-2.8}$}}
    \end{overpic}
    \hspace{0.3cm}
    \begin{overpic}[width=4cm]{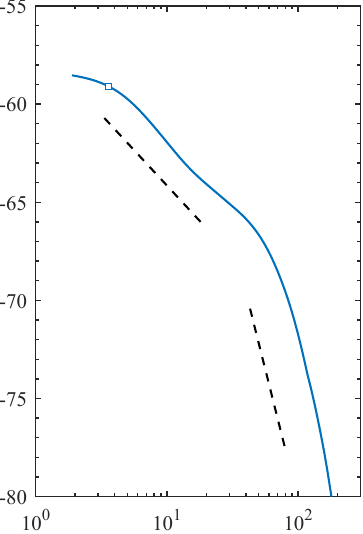}
        \put(-8,98){\color{black}{(b)}}
        \put(10,93){\color{black}\small{F3 (in ZPG)}}
        \put(15,57){\color{black}\footnotesize{$\sim\text{St}^{-0.7}$}}
        \put(32,28){\color{black}\footnotesize{$\sim\text{St}^{-2.8}$}}
    \end{overpic}
    \hspace{0.3cm}
    \begin{overpic}[width=4cm]{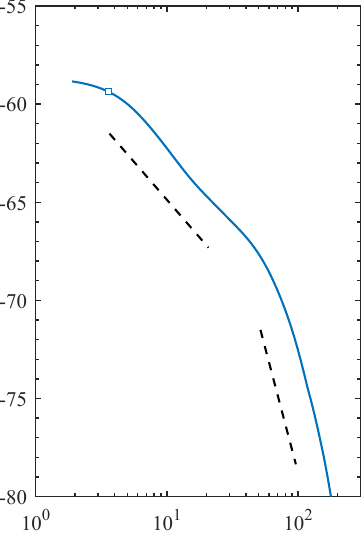}
        \put(-8,98){\color{black}{(c)}}
        \put(10,93){\color{black}\small{F4 (in ZPG)}}
        \put(15,57){\color{black}\footnotesize{$\sim\text{St}^{-0.7}$}}
        \put(34,23){\color{black}\footnotesize{$\sim\text{St}^{-2.8}$}}
    \end{overpic}
    \end{minipage}

    \vspace{0.5cm}
    
	\begin{minipage}[c]{1.0\linewidth}
    	\centering
    \begin{overpic}[width=4cm]{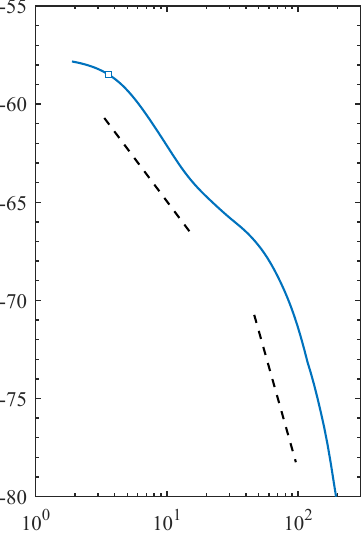}
        \put(-8,98){\color{black}{(d)}}
        \put(10,93){\color{black}\small{F5 (in FPG-1)}}
        \put(-10,20){{\color{black}\footnotesize{\rotatebox{90}{$10\log_{10}((\phi_{p^\prime p^\prime}/(\rho U_\infty^2)^2)(U_\infty/D))$}}}}
        \put(25,-5){{\color{black}\footnotesize{St = $fD/U_\infty$}}}
        \put(15,56){\color{black}\footnotesize{$\sim\text{St}^{-0.7}$}}
        \put(35,25){\color{black}\footnotesize{$\sim\text{St}^{-2.8}$}}
    \end{overpic}
    \hspace{0.3cm}
    \begin{overpic}[width=4cm]{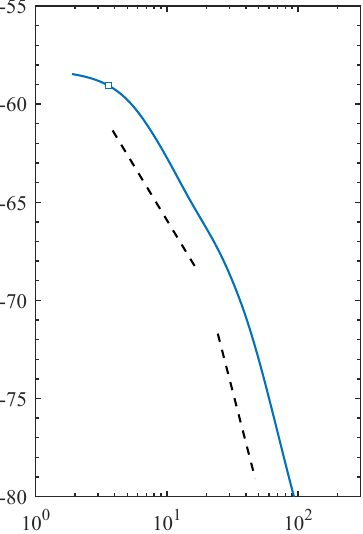}
        \put(-8,98){\color{black}{(e)}}
        \put(10,93){\color{black}\small{F6 (in APG-2)}}
        \put(25,-5){{\color{black}\footnotesize{St = $fD/U_\infty$}}}
        \put(15,56){\color{black}\footnotesize{$\sim\text{St}^{-1.2}$}}
        \put(25,25){\color{black}\footnotesize{$\sim\text{St}^{-2.8}$}}
    \end{overpic}
    \hspace{0.3cm}
    \begin{overpic}[width=4cm]{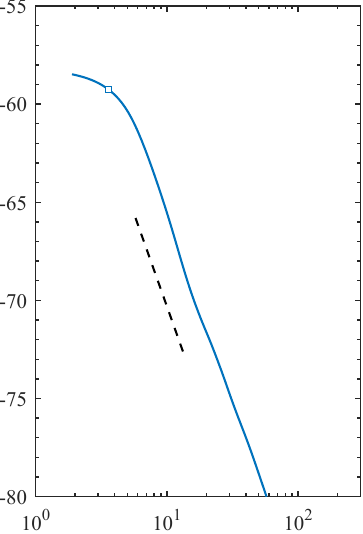}
        \put(-8,98){\color{black}{(f)}}
        \put(10,93){\color{black}\small{F7 (in FPG-2)}}
        \put(25,-5){{\color{black}\footnotesize{St = $fD/U_\infty$}}}
        \put(14,45){\color{black}\footnotesize{$\sim\text{St}^{-2}$}}
    \end{overpic}
    \end{minipage}
    \vspace{0.5cm}
    \caption{Power spectral density of fluctuating pressure measured at six sensors (F1-F6) along the SUBOFF hull under straight-ahead conditions, covering Reynolds numbers from $5.6 \times 10^6$ to $1.4 \times 10^7$. Subfigures (a) F2, (b) F5, (c) F6, and (d) F7 illustrate the PSD in dB/Hz relative to a reference pressure of 20 $\mu$Pa.}
    \label{fig:scaling_Fluc_30_0deg}
\end{figure*}
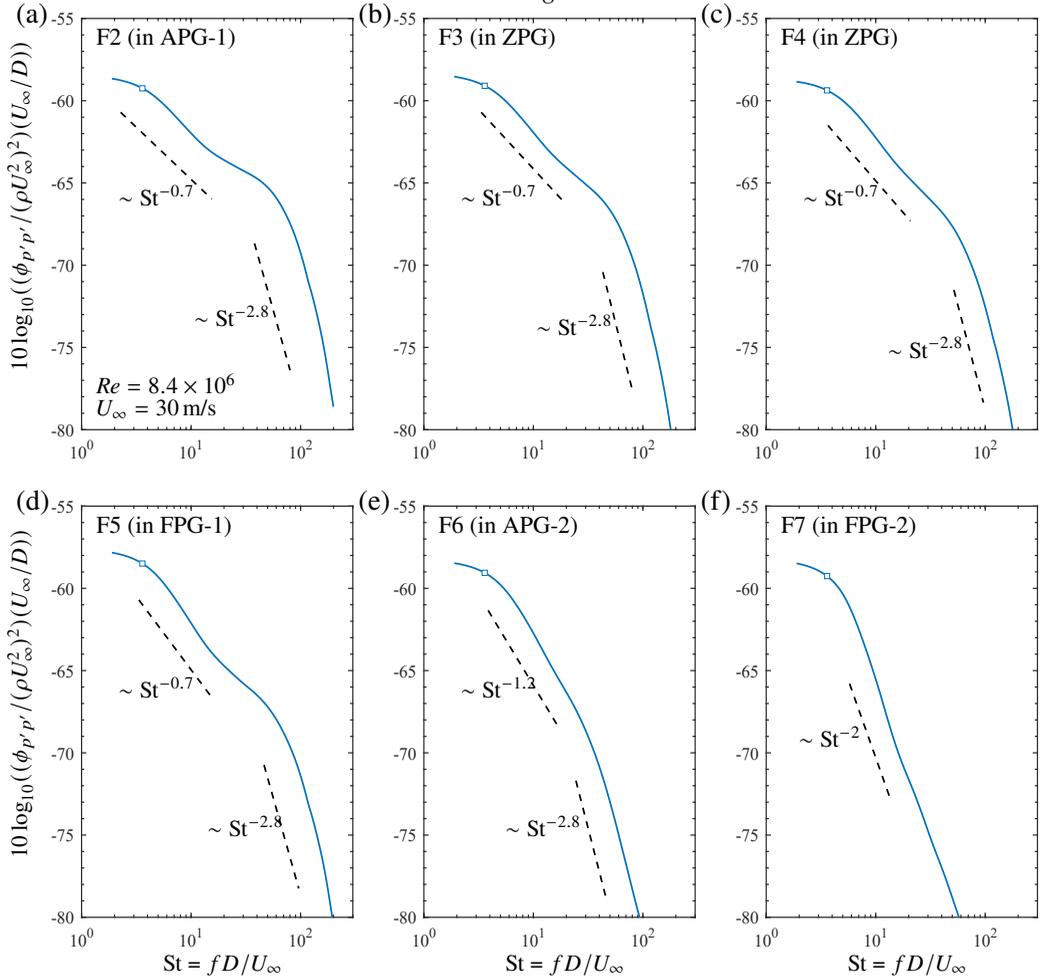

To analyze the scaling characteristics of the turbulent pressure field, the wall-pressure spectra are normalized using outer variables. The non-dimensional power spectral density is defined as $10\log_{10}(\phi_{p^\prime p^\prime}/(\rho U_\infty^2)^2)(U_\infty/D))$, with the frequency normalized as the Strouhal number, $\text{St} = fD/U_\infty$, where $D$ is the maximum hull diameter. This analysis uses the data at $U_\infty = 30 \, \text{m/s}$ ($\Rey = 8.4 \times 10^6$) as it captures a wide frequency range, and as noted previously, the scaling behaviors are similar across the tested Reynolds numbers. Figure~\ref{fig:scaling_Fluc_30_0deg} presents these non-dimensional spectra for sensors F2 through F7. The plots for each sensor demonstrate that this outer scaling successfully captures the spectral data, particularly in the mid- and high-frequency ranges. In the zero-pressure-gradient (ZPG) region, represented by sensors F3 (figure~\ref{fig:scaling_Fluc_30_0deg}(b)) and F4 (figure~\ref{fig:scaling_Fluc_30_0deg}(c)), the spectra exhibit a clear two-slope behavior. The mid-frequency range shows a shallow decay proportional to $\sim\text{St}^{-0.7}$, followed by a steeper roll-off in the high-frequency range proportional to $\sim\text{St}^{-2.8}$. This $\sim\text{St}^{-0.7}$ slope in the ZPG region is consistent with observations from several smooth-wall turbulent boundary layer studies, which also report mid-frequency slopes between -0.7 and -0.8. The sensors in the mild pressure-gradient regions near the fore and mid-body show similar characteristics. Sensor F2 (in APG-1, figure~\ref{fig:scaling_Fluc_30_0deg}(a)) and F5 (in FPG-1, figure~\ref{fig:scaling_Fluc_30_0deg}(d)) both display spectral shapes and slopes ($\sim\text{St}^{-0.7}$ and $\sim\text{St}^{-2.8}$) that are nearly identical to the ZPG cases. This indicates that the moderate pressure gradients in these areas do not significantly alter the scaled energy distribution. However, significant deviations are observed in the stern region where pressure gradients are stronger and flow history effects become dominant. At sensor F6 (in APG-2, figure~\ref{fig:scaling_Fluc_30_0deg}(e)), the mid-frequency slope steepens to $\sim\text{St}^{-1.2}$, indicating a different energy cascade mechanism as the boundary layer approaches the tail under an adverse pressure gradient. The high-frequency slope remains $\sim\text{St}^{-2.8}$. At sensor F7 (in FPG-2, figure~\ref{fig:scaling_Fluc_30_0deg}(f)), the spectrum is most distinct, characterized by a more uniform decay of $\sim\text{St}^{-2}$ across a wide frequency range. This change reflects the complex flow dynamics at the extreme aft-end, including pressure recovery and interaction with the developing wake. It is noteworthy that the observed high-frequency slope of $\sim\text{St}^{-2.8}$ differs from the theoretical $f^{-5}$ viscous roll-off reported in studies using viscous or wall-wake scaling, such as the $f^{-5}$ decay noted by \citet{Balantrapu2023APG} in a strong APG flow. This suggests that the high-frequency range captured in this study, when non-dimensionalized with outer variables, is not yet in the purely viscous subrange and its scaling is still strongly influenced by the outer-flow parameters $D$ and $U_\infty$.

\subsection{Effects of maneuvering conditions on wall-pressure spectra}\label{Maneuvering_Conditions}
\subsubsection{Effect of the yaw angle}\label{Yaw_Angle}
\begin{figure*} 
    \centering
    \begin{minipage}[c]{1.0\linewidth}
    	\centering
    \begin{overpic}[width=11cm]{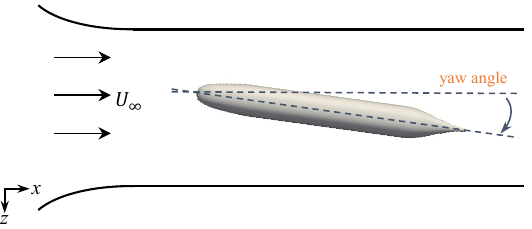}
        \put(-2,40){\color{black}{(a)}}
    \end{overpic}
    \end{minipage}
    
    \vspace{0.1cm}
	\begin{minipage}[c]{1.0\linewidth}
    	\centering
    \begin{overpic}[width=4cm]{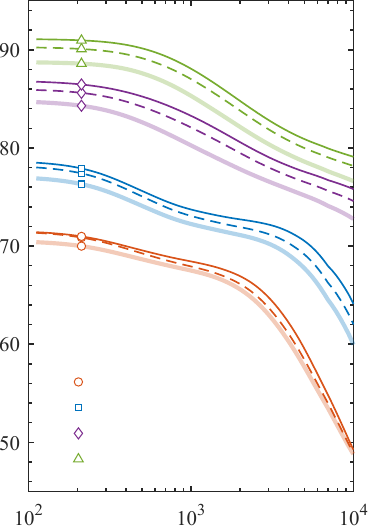}
        \put(-5,98){\color{black}{(b)}}
        \put(10,95){\color{black}\footnotesize{F2 (in APG-1)}}
        \put(18,35){\color{black}\footnotesize{dB re 20 $\mu$Pa}}
        \put(18,26){\color{black}\footnotesize{$\Rey = 5.6 \times 10^6$}}
        \put(18,21){\color{black}\footnotesize{$\Rey = 8.4 \times 10^6$}}
        \put(18,16){\color{black}\footnotesize{$\Rey = 1.2 \times 10^7$}}
        \put(18,11.5){\color{black}\footnotesize{$\Rey = 1.4 \times 10^7$}}
        \put(-10,20){\small{\color{black}{\rotatebox{90}{$10 \log_{10}(\phi_{p^\prime p^\prime}/p^2_\text{ref})\, \text{(dB/Hz)}$}}}}
    \end{overpic}
    \hspace{0.1cm}
    \begin{overpic}[width=4cm]{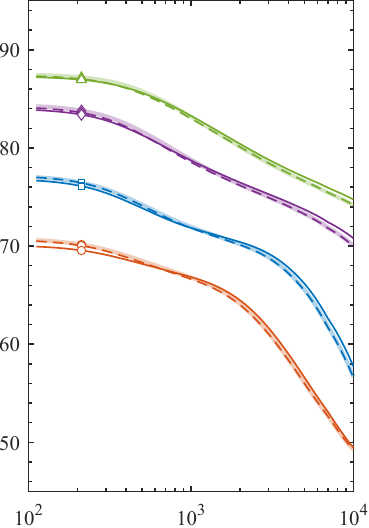}
        \put(-5,98){\color{black}{(c)}}
        \put(10,93){\color{black}\footnotesize{F3 (in ZPG)}}
    \end{overpic}
    \hspace{0.1cm}
    \begin{overpic}[width=4cm]{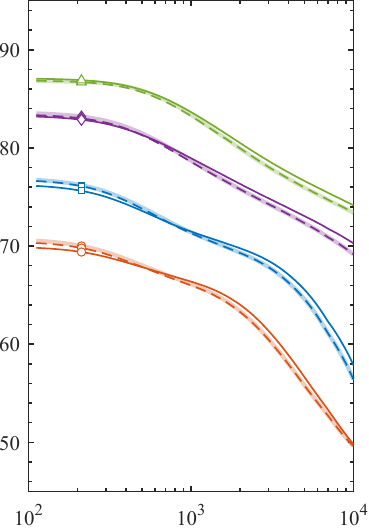}
        \put(-5,98){\color{black}{(d)}}
        \put(10,93){\color{black}\footnotesize{F4 (in ZPG)}}
    \end{overpic}
    \end{minipage}

    \vspace{0.1cm}
    
	\begin{minipage}[c]{1.0\linewidth}
    	\centering
    \begin{overpic}[width=4cm]{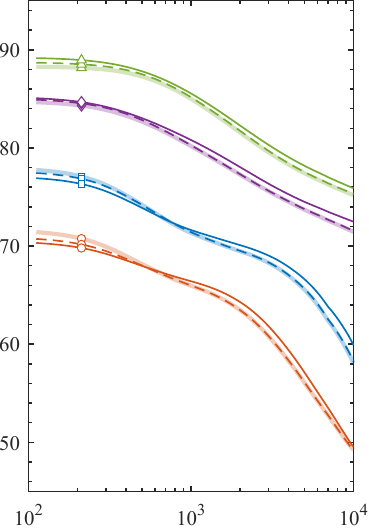}
        \put(-5,98){\color{black}{(e)}}
        \put(10,93){\color{black}\footnotesize{F5 (in FPG-1)}}
        \put(-10,20){\small{\color{black}{\rotatebox{90}{$10 \log_{10}(\phi_{p^\prime p^\prime}/p^2_\text{ref})\, \text{(dB/Hz)}$}}}}
        \put(29,-5){\small{\color{black}{$f$ (Hz)}}}
    \end{overpic}
    \hspace{0.1cm}
    \begin{overpic}[width=4cm]{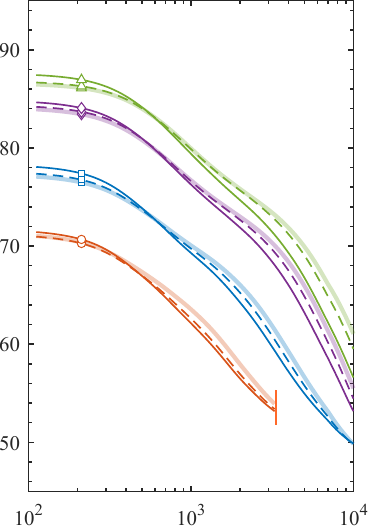}
        \put(-5,98){\color{black}{(f)}}
        \put(10,93){\color{black}\footnotesize{F6 (in APG-2)}}
        \put(29,-5){\small{\color{black}{$f$ (Hz)}}}
    \end{overpic}
    \hspace{0.1cm}
    \begin{overpic}[width=4cm]{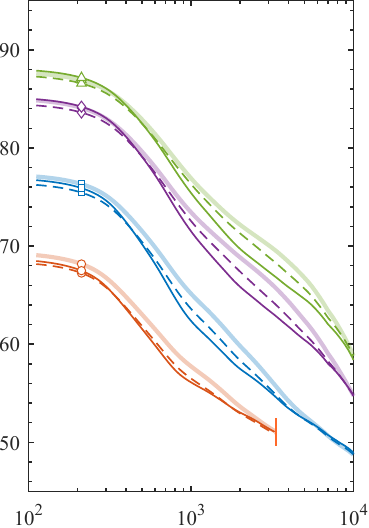}
        \put(-5,98){\color{black}{(g)}}
        \put(10,93){\color{black}\footnotesize{F7 (in FPG-2)}}
        \put(29,-5){\small{\color{black}{$f$ (Hz)}}}
    \end{overpic}
    \end{minipage}
    \vspace{0.1cm}
    \caption{Comparison of the PSD of pressure fluctuations under yaw angles (Yaw $3^\circ$ by dashed line, Yaw $6^\circ$ by solid line) with the straight ahead condition (light solid line), covering Reynolds numbers from $5.6 \times 10^6$ to $1.4 \times 10^7$. (a) Schematic of the yaw angle maneuver. Subfigures (b) F2, (c) F3, (d) F4, (e) F5, (f) F6, and (g) F7 illustrate the PSD in dB/Hz relative to a reference pressure of 20 $\mu$Pa.}
    \label{fig:effect_yaw_angle}
\end{figure*}
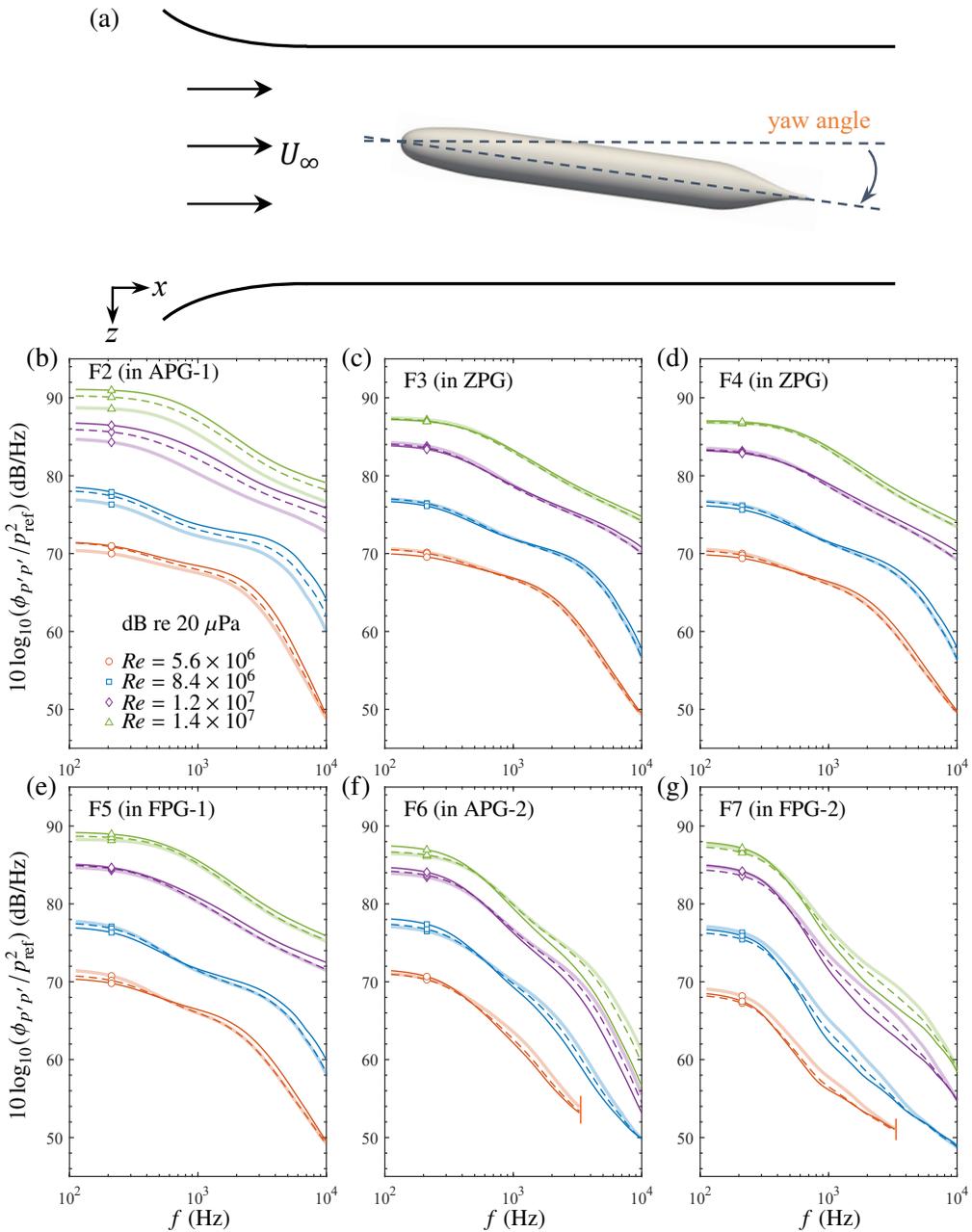


This section analyzes the response of wall-pressure fluctuations under horizontal turning (yaw) maneuvers. Figure~\ref{fig:effect_yaw_angle} presents the power spectral density at yaw angles of $3^\circ$ (dashed line) and $6^\circ$ (solid line), compared against the $0^\circ$ straight-ahead condition (light solid line) across all four Reynolds numbers. The most significant and consistent trend observed is that the introduction of a yaw angle systematically increases the wall-pressure fluctuation levels across the entire frequency spectrum. This effect is monotonic with the yaw angle: the PSD levels at $6^\circ$ are uniformly higher than those at $3^\circ$, which are, in turn, significantly higher than the $0^\circ$ baseline. This phenomenon is observed at all sensor locations and across the full range of Reynolds numbers tested ($\Rey = 5.6 \times 10^6$ to $1.4 \times 10^7$). This amplification is physically attributed to the three-dimensional crossflow induced by the yawed condition. As the hull operates at an angle of attack to the incoming flow, the flow on the top meridian line (leeward side) is subjected to separation, leading to the formation of strong, unsteady crossflow vortices. This highly three-dimensional and separated flow is a much more strong source of pressure fluctuations than the attached boundary layer of the straight-ahead case. A detailed examination of the sensor locations reveals:

\begin{enumerate}[label=(\roman*)]
\item Forebody (F2): In the forward adverse pressure gradient region (F2 in APG-1), the yaw angle has a clear and systematic effect. The PSD levels consistently increase with the yaw angle, and this amplification becomes more obvious at higher Reynolds numbers. As seen in figure~\ref{fig:effect_yaw_angle}(b), at the maximum $\Rey = 1.4 \times 10^7$ (green lines), the $6^\circ$ yaw angle results in a uniform PSD increase of approximately 2 dB across the entire spectrum compared to the straight-ahead condition. This indicates that the initial crossflow induced by yaw at the nose disturb the boundary layer in this adverse pressure gradient region.

\item Parallel Mid-body (F3, F4, F5): In stark contrast, the sensors located in the parallel mid-body show a remarkable insensitivity to yaw. For F3 (in ZPG), F4 (in ZPG), and F5 (in FPG-1 just behind the ZPG region), the PSD curves for the $0^\circ$, $3^\circ$, and $6^\circ$ cases at any given Reynolds number nearly collapse onto a single line (figure~\ref{fig:effect_yaw_angle}(c-e)) with less than 1 dB difference. This suggests that in this region of zero or favorable pressure gradient, the flow remains largely attached, and the 3D crossflow effects are not yet strong enough to significantly alter the wall-pressure signature.

\item Aft-body (F6, F7): The most complex response occurs at the aft-body. Here, the yaw angle's primary influence is on the high-frequency content ($f >  1000$ Hz). Contrary to the forebody, increasing the yaw angle leads to a distinct `suppression' of high-frequency pressure fluctuations (figure~\ref{fig:effect_yaw_angle}(f, g)). This suppression is monotonic: the $6^\circ$ yaw angle (solid line) exhibits the lowest high-frequency energy, followed by the $3^\circ$ angle (dashed line), while the straight-ahead condition (light solid line) retains the highest high-frequency energy. This phenomenon is attributed to the formation of large-scale, 3D crossflow separation vortices on the leeward side of the tail under yaw. As these large, stable vortices form, they effectively lift the primary shear layer-and its associated small-scale, high-frequency turbulent structures-away from the hull surface. The sensors at F6 and F7 are thus located within a region of relatively slower, separated flow, shielded from the high-frequency turbulence generation, which results in the observed high-frequency attenuation.
\end{enumerate}

Furthermore, a comparison across Reynolds numbers suggests that the disparity in PSD between the yawed and straight-ahead conditions widens at higher $\Rey$. This implies that the flow becomes more sensitive to three-dimensional separation and its associated unsteady effects as the Reynolds number increases.

\subsubsection{Effect of the pitch angle}\label{pitch_angle}
\begin{figure*} 
    \centering
    \begin{minipage}[c]{1.0\linewidth}
    	\centering
    \begin{overpic}[width=11cm]{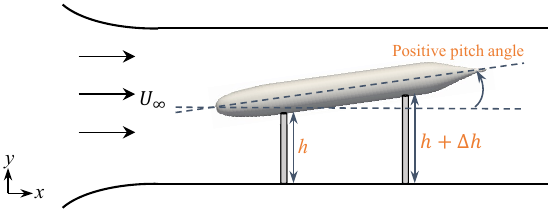}
        \put(5,36){\color{black}{(a)}}
    \end{overpic}
    \end{minipage}
    
    \vspace{0.1cm}
	\begin{minipage}[c]{1.0\linewidth}
    	\centering
    \begin{overpic}[width=4cm]{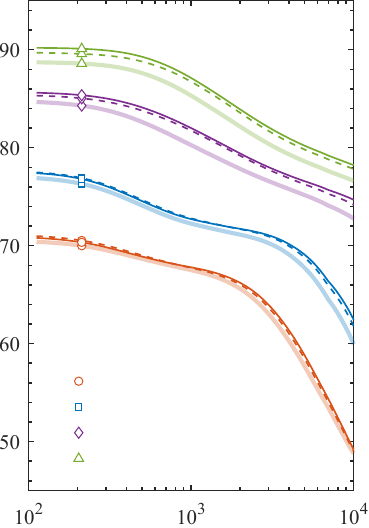}
        \put(-5,98){\color{black}{(b)}}
        \put(10,95){\color{black}\footnotesize{F2 (in APG-1)}}
        \put(18,35){\color{black}\footnotesize{dB re 20 $\mu$Pa}}
        \put(18,26){\color{black}\footnotesize{$\Rey = 5.6 \times 10^6$}}
        \put(18,21){\color{black}\footnotesize{$\Rey = 8.4 \times 10^6$}}
        \put(18,16){\color{black}\footnotesize{$\Rey = 1.2 \times 10^7$}}
        \put(18,11.5){\color{black}\footnotesize{$\Rey = 1.4 \times 10^7$}}
        \put(-10,20){\small{\color{black}{\rotatebox{90}{$10 \log_{10}(\phi_{p^\prime p^\prime}/p^2_\text{ref})\, \text{(dB/Hz)}$}}}}
    \end{overpic}
    \hspace{0.1cm}
    \begin{overpic}[width=4cm]{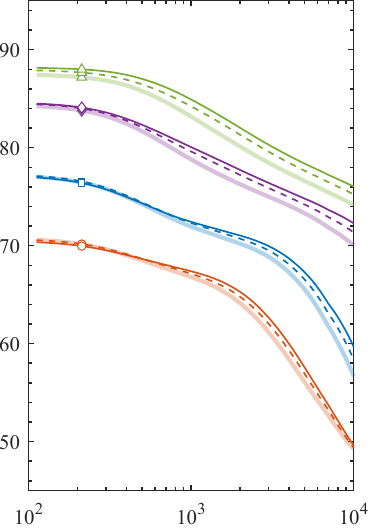}
        \put(-5,98){\color{black}{(c)}}
        \put(10,93){\color{black}\footnotesize{F3 (in ZPG)}}
    \end{overpic}
    \hspace{0.1cm}
    \begin{overpic}[width=4cm]{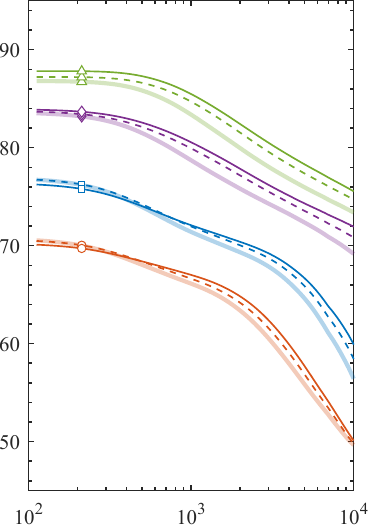}
        \put(-5,98){\color{black}{(d)}}
        \put(10,93){\color{black}\footnotesize{F4 (in ZPG)}}
    \end{overpic}
    \end{minipage}

    \vspace{0.1cm}
    
	\begin{minipage}[c]{1.0\linewidth}
    	\centering
    \begin{overpic}[width=4cm]{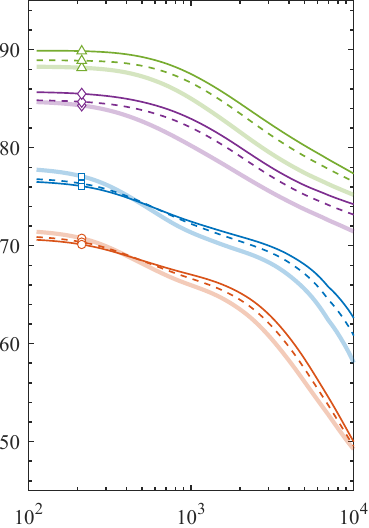}
        \put(-5,98){\color{black}{(e)}}
        \put(10,93){\color{black}\footnotesize{F5 (in FPG-1)}}
        \put(-10,20){\small{\color{black}{\rotatebox{90}{$10 \log_{10}(\phi_{p^\prime p^\prime}/p^2_\text{ref})\, \text{(dB/Hz)}$}}}}
        \put(29,-5){\small{\color{black}{$f$ (Hz)}}}
    \end{overpic}
    \hspace{0.1cm}
    \begin{overpic}[width=4cm]{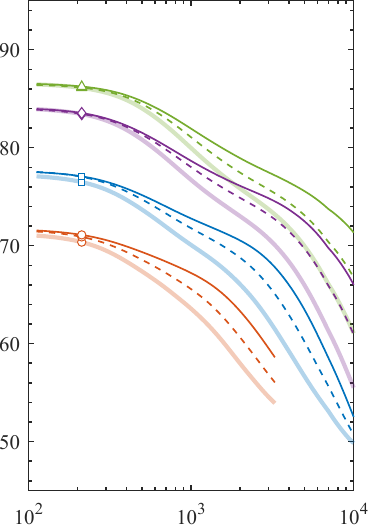}
        \put(-5,98){\color{black}{(f)}}
        \put(10,93){\color{black}\footnotesize{F6 (in APG-2)}}
        \put(29,-5){\small{\color{black}{$f$ (Hz)}}}
    \end{overpic}
    \hspace{0.1cm}
    \begin{overpic}[width=4cm]{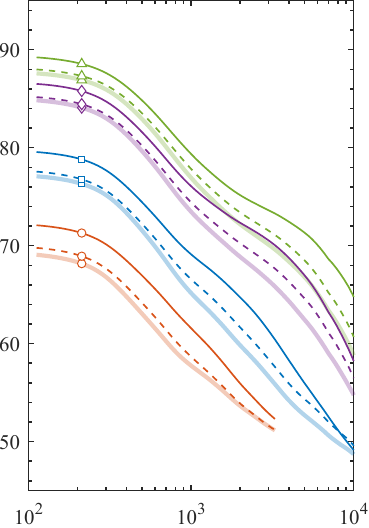}
        \put(-5,98){\color{black}{(g)}}
        \put(10,93){\color{black}\footnotesize{F7 (in FPG-2)}}
        \put(29,-5){\small{\color{black}{$f$ (Hz)}}}
    \end{overpic}
    \end{minipage}
    \vspace{0.1cm}
    \caption{Comparison of the PSD of pressure fluctuations under positive pitch angles (Pitch $+3^\circ$ by dashed line, Pitch $+6^\circ$ by solid line) with the straight ahead condition (light solid line), covering Reynolds numbers from $5.6 \times 10^6$ to $1.4 \times 10^7$. (a) Schematic of the positive pitch angle (bow-down) maneuver. Subfigures (b) F2, (c) F3, (d) F4, (e) F5, (f) F6, and (g) F7 illustrate the PSD in dB/Hz relative to a reference pressure of 20 $\mu$Pa.}
    \label{fig:effect_pitch_angle_positive}
\end{figure*}
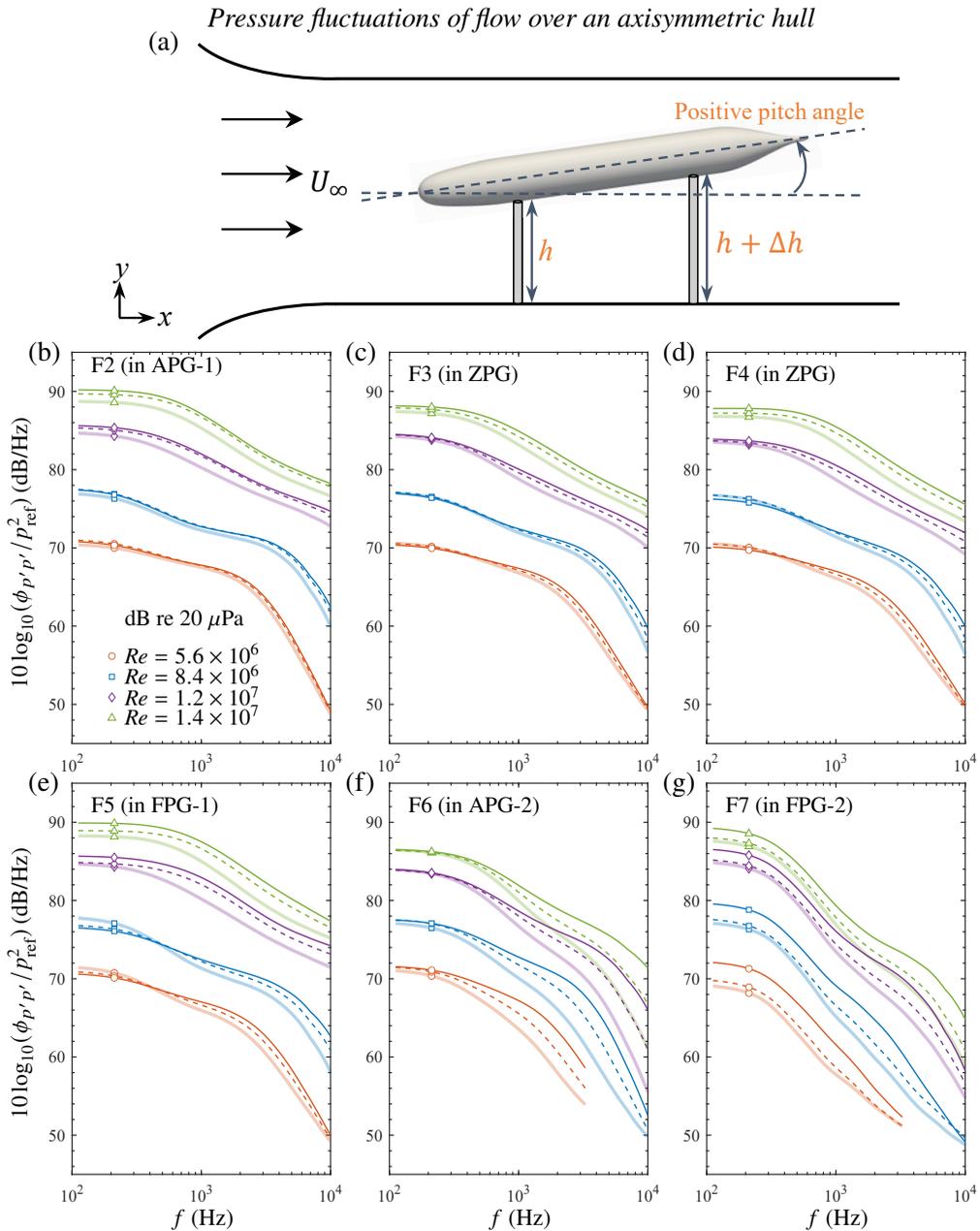

\begin{figure*} 
    \centering
    \begin{minipage}[c]{1.0\linewidth}
    	\centering
    \begin{overpic}[width=11cm]{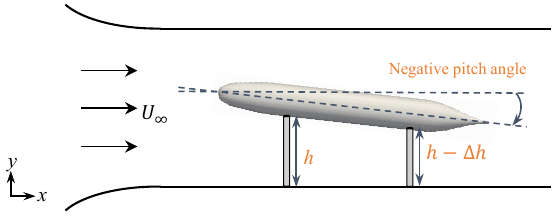}
        \put(5,36){\color{black}{(a)}}
    \end{overpic}
    \end{minipage}
    
    \vspace{0.1cm}
    
	\begin{minipage}[c]{1.0\linewidth}
    	\centering
    \begin{overpic}[width=4cm]{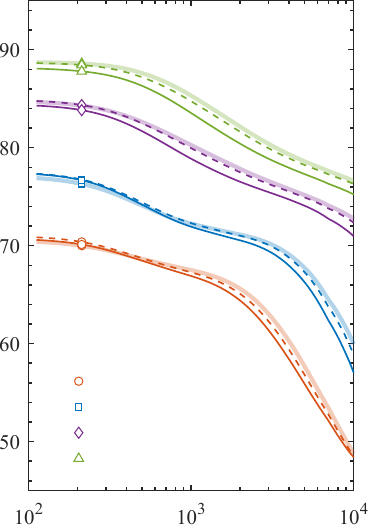}
        \put(-5,98){\color{black}{(b)}}
        \put(10,95){\color{black}\footnotesize{F2 (in APG-1)}}
        \put(18,35){\color{black}\footnotesize{dB re 20 $\mu$Pa}}
        \put(18,26){\color{black}\footnotesize{$\Rey = 5.6 \times 10^6$}}
        \put(18,21){\color{black}\footnotesize{$\Rey = 8.4 \times 10^6$}}
        \put(18,16){\color{black}\footnotesize{$\Rey = 1.2 \times 10^7$}}
        \put(18,11.5){\color{black}\footnotesize{$\Rey = 1.4 \times 10^7$}}
        \put(-10,20){\small{\color{black}{\rotatebox{90}{$10 \log_{10}(\phi_{p^\prime p^\prime}/p^2_\text{ref})\, \text{(dB/Hz)}$}}}}
    \end{overpic}
    \hspace{0.1cm}
    \begin{overpic}[width=4cm]{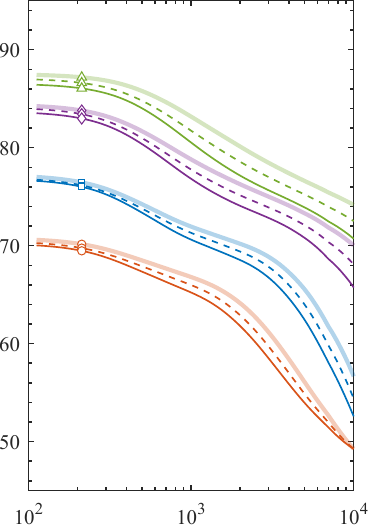}
        \put(-5,98){\color{black}{(c)}}
        \put(10,93){\color{black}\footnotesize{F3 (in ZPG)}}
    \end{overpic}
    \hspace{0.1cm}
    \begin{overpic}[width=4cm]{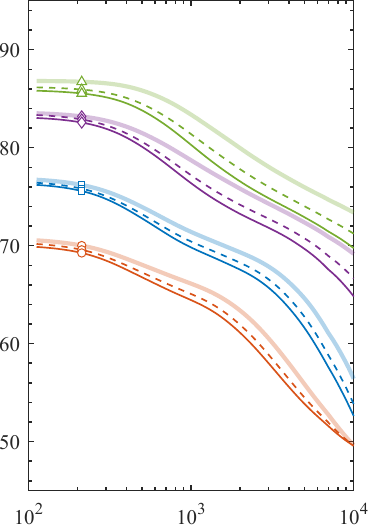}
        \put(-5,98){\color{black}{(d)}}
        \put(10,93){\color{black}\footnotesize{F4 (in ZPG)}}
    \end{overpic}
    \end{minipage}

    \vspace{0.1cm}
    
	\begin{minipage}[c]{1.0\linewidth}
    	\centering
    \begin{overpic}[width=4cm]{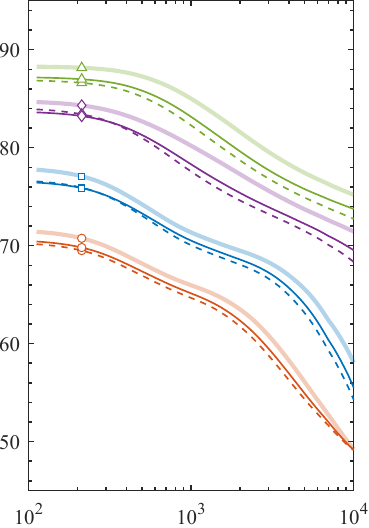}
        \put(-5,98){\color{black}{(e)}}
        \put(10,93){\color{black}\footnotesize{F5 (in FPG-1)}}
        \put(-10,20){\small{\color{black}{\rotatebox{90}{$10 \log_{10}(\phi_{p^\prime p^\prime}/p^2_\text{ref})\, \text{(dB/Hz)}$}}}}
        \put(29,-5){\small{\color{black}{$f$ (Hz)}}}
    \end{overpic}
    \hspace{0.1cm}
    \begin{overpic}[width=4cm]{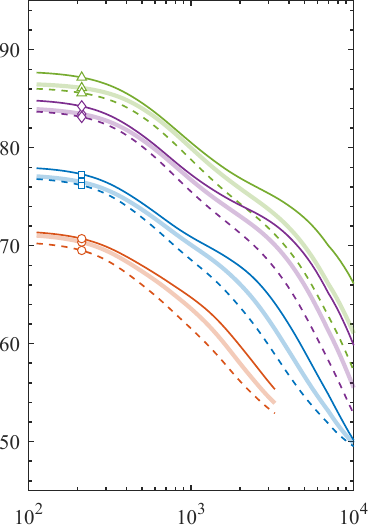}
        \put(-5,98){\color{black}{(f)}}
        \put(10,93){\color{black}\footnotesize{F6 (in APG-2)}}
        \put(29,-5){\small{\color{black}{$f$ (Hz)}}}
    \end{overpic}
    \hspace{0.1cm}
    \begin{overpic}[width=4cm]{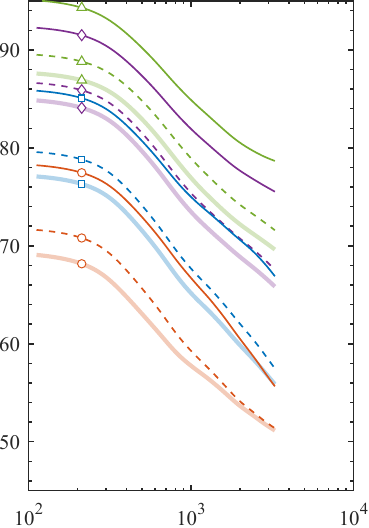}
        \put(-5,98){\color{black}{(g)}}
        \put(35,93){\color{black}\footnotesize{F7 (in FPG-2)}}
        \put(29,-5){\small{\color{black}{$f$ (Hz)}}}
    \end{overpic}
    \end{minipage}
    \vspace{0.1cm}
    \caption{Comparison of the PSD of pressure fluctuations under negative pitch angles (Pitch $-3^\circ$ by dashed line, Pitch $-6^\circ$ by solid line) with the straight ahead condition (light solid line), covering Reynolds numbers from $5.6 \times 10^6$ to $1.4 \times 10^7$. (a) Schematic of the negative pitch angle (bow-up) maneuver. Subfigures (b) F2, (c) F3, (d) F4, (e) F5, (f) F6, and (g) F7 illustrate the PSD in dB/Hz relative to a reference pressure of 20 $\mu$Pa.}
    \label{fig:effect_pitch_angle_negative}
\end{figure*}

The influence of pitch (heave) maneuvers on wall-pressure fluctuations is analyzed, revealing highly distinct responses depending on whether the sensor line is on the windward or leeward side. First, the positive pitch angle (bow-down) condition is addressed, as depicted in figure~\ref{fig:effect_pitch_angle_positive}(a). In this maneuver, the top meridian line, where the sensors are located, becomes the windward side of the hull. The flow is forced to accelerate over this surface. This generally leads to a thinner, more energetic boundary layer with intensified shear production, resulting in an amplification of wall-pressure fluctuations. However, the specific response varies significantly along the hull:

\begin{enumerate}[label=(\roman*)]
\item Forebody (F2): At the forebody (F2 in APG-1), a clear broadband amplification is observed (figure~\ref{fig:effect_pitch_angle_positive}(b)). The PSD curves shift upwards monotonically with the pitch angle across all frequencies. This effect is magnified at higher Reynolds numbers, suggesting that the thinner, high-$\Rey$ boundary layer is more sensitive to the increased local shear and turbulence production induced by the impinging flow.

\item Parallel mid-body (F3, F4, F5): A more complex, frequency-dependent response occurs in the parallel mid-body (figure~\ref{fig:effect_pitch_angle_positive}(c-e)). In the mid-to-high frequency range ($f > 300$ Hz), the PSD is clearly amplified, and this amplification increases with both pitch angle and Reynolds number. Conversely, in the low-frequency range ($f < 300$ Hz), the PSD levels are slightly suppressed below the straight-ahead baseline. This complex behavior suggests that while the windward orientation enhances the small-scale, high-frequency turbulent structures (due to increased mean shear), it simultaneously stabilizes or displaces the large-scale, low-frequency coherent structures in this ZPG/FPG-1 region.

\item Aft-body (F6, F7): The aft-body sensors show the most dramatic amplification. At F6 (in APG-2, figure~\ref{fig:effect_pitch_angle_positive}(f)), the behavior resembles the mid-body but with a much more pronounced high-frequency amplification. At $\Rey = 1.4 \times 10^7$ (green lines), the $+6^\circ$ pitch angle results in a $\sim 5$ dB increase at high frequencies. As you hypothesized, the flow likely remains attached due to the energizing effect of the windward flow, which counteracts the natural adverse pressure gradient and results in intense shear. Finally, at sensor F7 (in FPG-2, figure~\ref{fig:effect_pitch_angle_positive}(g)), the response is a strong, broadband amplification across all frequencies, similar to F2. The separation between curves is distinct even at the lowest $\Rey$, with the $+6^\circ$ case showing an increase of $\sim 2$ dB. This indicates a uniformly more energetic turbulent field at the extreme stern.
\end{enumerate}

Conversely, the negative pitch angle (bow-up) condition, illustrated in figure~\ref{fig:effect_pitch_angle_negative}(a), presents a far more complex flow phenomenon. In this maneuver, the top meridian line, where the sensors are located, becomes the leeward side of the hull. This configuration is highly susceptible to three-dimensional flow separation.

\begin{enumerate}[label=(\roman*)]
\item Fore and mid-body (F2, F3, F4, F5): For sensors F2 (APG-1), F3 (ZPG), F4 (ZPG), and F5 (FPG-1), the results align with the expected physics of leeward-side separation. A large-scale, stable separation bubble forms on the leeward side, shielding the sensors from the high-energy turbulence in the outer shear layer.

This shielding effect is evident as the pitched cases ($-3^\circ$ dashed line and $-6^\circ$ solid line) show a distinct suppression of the PSD relative to the $0^\circ$ straight-ahead condition (light solid line). This suppression is most pronounced in the mid-to-high frequency range for F2, F3, and F4 (figure~\ref{fig:effect_pitch_angle_negative}(b-d)), as the small-scale, high-frequency turbulent structures are lifted away from the wall. By the time the flow reaches F5 (figure~\ref{fig:effect_pitch_angle_negative}(e)), the separation is likely fully established, leading to a more consistent broadband suppression. The fact that this suppression becomes more pronounced at higher Reynolds numbers is consistent with the separation bubble becoming more defined.

\item Aft-body (F7, in FPG-2): The complete reversal
sensor F7 ($x/L = 0.956$), located near the tail tip, shows a behavior that is the complete opposite of the mid-body: a massive, monotonic amplification of the PSD (figure~\ref{fig:effect_pitch_angle_negative}(g)). Both the $-3^\circ$ and $-6^\circ$ cases are significantly higher than the $0^\circ$ baseline. This amplification increases dramatically with both pitch angle and Reynolds number, reaching a difference of nearly 8 dB at the highest $\Rey$ and $-6^\circ$ pitch. This dramatic reversal indicates that F7 is not in a separated region. Instead, it is likely being subjected to the direct influence of 3D crossflow vortices. As the flow separates from the leeward side, it rolls up into vortices. The severe curvature of the hull at the extreme tail (in FPG-2) likely forces these highly unsteady vortex cores to impinge upon the hull surface at this location. This direct vortex-surface interaction act as a strong source of pressure fluctuations, which would explain the intense broadband amplification observed.

\item Aft-body (F6, in APG-2): The transitional phenomenon
sensor F6 ($x/L = 0.871$) exhibits the complex, non-monotonic behavior you identified (figure~\ref{fig:effect_pitch_angle_negative}(f)): the $-3^\circ$ case (dashed line) shows suppression, while the $-6^\circ$ case (solid line) shows amplification (both relative to $0^\circ$), primarily in the mid-to-high frequencies. F6 is located in the highly sensitive APG-2 region, physically positioned between the stable separated flow of the mid-body (F5) and the intense vortex-dominated region at the tail tip (F7). It is at the boundary of two competing flow regimes: At $-3^\circ$ (Mild Pitch): The leeward separation bubble, which dominates F5, is likely still the primary flow feature at F6. The flow is separated, resulting in the observed suppression. At $-6^\circ$ (strong pitch): The angle of attack is doubled, creating a much stronger 3D crossflow and more powerful tail vortices. The influence of this energetic vortex system now extends upstream to F6. The flow at F6 transitions from a stable separated state (at -3°) to a highly unsteady, vortex-influenced, or intermittently reattaching state. This new, more energetic regime causes the PSD to amplify, beginning to mirror the behavior seen at F7.
\end{enumerate}

In summary, the results for F6 and F7 capture the complex 3D physics of leeward-side flow: a stable separation on the mid-body (F2-F5, causing suppression) that transitions into a highly energetic, vortex-dominated flow at the tail (F7, causing amplification), with F6 capturing the precise transitional boundary between these two regimes.

\section{Conclusions}\label{Concluding}

This paper has presented a comprehensive experimental investigation into the wall-pressure fluctuations on an axisymmetric (SUBOFF) hull. The primary innovation of this study is the establishment of the first, to our knowledge, high-fidelity database of wall-pressure fluctuations at high Reynolds numbers ($\Rey = 5.6 \times 10^6$ to $1.4 \times 10^7$) that systematically includes both straight-ahead (multi-speed) and complex maneuvering (yaw and pitch) conditions. Through rigorous signal correction and validation, this dataset provides an important benchmark for advanced turbulence models and hydroacoustic prediction tools, particularly for off-design (maneuvering) cases.

The systematic analysis of the straight-ahead condition revealed key physical insights into the baseline flow. While spectral levels uniformly increase with $\Rey$, the influence of pressure gradients was found to be complex. The forward APG region (F2) exhibited spectral amplification at high $\Rey$, while the stern region (F6 and F7) showed a distinct suppression of mid-to-high frequency content relative to the mid-body. This highlights the dominance of flow history and hull curvature on small-scale turbulence at the stern. Consequently, standard spectral scaling based on outer variables (e.g., $\sim\text{St}^{-0.7}$), which held for the fore and mid-body, breaks down in the aft-body (F6: $\sim\text{St}^{-1.2}$, F7: $\sim\text{St}^{-2}$), confirming the tail flow is governed by local, non-equilibrium physics.

The investigation into maneuvering conditions demonstrated that hull attitude fundamentally alters the pressure field, with the response dictated by the sensor's position relative to the windward or leeward side. For yaw maneuvers, the leeward top meridian showed insensitivity in the mid-body (F3-F5) but a distinct high-frequency suppression at the tail (F6, F7), attributed to stable crossflow vortices lifting the small-scale turbulent structures. Conversely, for positive pitch (bow-down) maneuvers, the sensors on the windward side were energized, inducing strong broadband amplification at the fore and aft (F2, F7).

One of the most important findings of this work is the discovery of the complex, transitional physics during the negative pitch (bow-up) maneuver. This leeward condition revealed a dramatic spatial bifurcation: while the fore and mid-body (F2-F5) showed the expected spectral suppression due to a stable separation bubble, the extreme tail (F7) exhibited a massive, monotonic amplification (up to 8 dB). Concurrently, sensor F6 (just upstream) captured a non-monotonic, transitional behavior (suppression at -3°, amplification at -6°). This is not an experimental artifact but rather the first clear experimental evidence of a complex spatial flow transition: from a stable leeward separation bubble on the mid-body to a highly energetic region at the tail tip dominated by the direct impingement of 3D crossflow vortices.

This foundational research, in conjunction with the benchmark dataset it provides, establishes the foundation for future studies to explore a range of potential research topics: (i) CFD Validation: Rigorously validating RANS, DES, and LES models for flow nosie prediction, particularly their ability to capture the complex pressure fluctuations. (ii) Appendage Effects: Extending the experimental test model to include appendages (such as the sail and fins), which are known to be dominant local noise sources, and studying their interaction with the hull's maneuvering flow field. (iii) Hydroacoustic Modeling: Using the measured PSDs as direct input for numerical (BEM/FEM) or statistical (SEA) models to predict the far-field radiated noise. Ultimately, the insights and data generated from this study provide a critical step toward the physics-based prediction and reduction of flow noise, enabling the design of quieter and more effective underwater vehicles.

\begin{bmhead}[Acknowledgements] 
This work was supported in part by National Natural Science Foundation of China (Grant No. 52522113), the Shanghai Pilot Program for Basic Research of Shanghai Jiao Tong University (No. 21TQ1400202), and the Fundamental Research Funds for the Central Universities.   
\end{bmhead}

\begin{bmhead}[Declaration of Interests] 
The authors report no conflict of interest.
\end{bmhead}

\begin{bmhead}[Data availability statement] 
The data that support the findings of this study are available from the corresponding author upon reasonable request.
\end{bmhead}

\begin{bmhead}[Author ORCIDs.] 
\\
\noindent Peng Jiang, \href{https://orcid.org/0000-0002-9072-8307}{https://orcid.org/0000-0002-9072-8307};\\
Bin Xie, \href{https://orcid.org/0000-0002-4218-2442}{https://orcid.org/0000-0002-4218-2442};\\
Shijun Liao, \href{https://orcid.org/0000-0002-2372-9502}{https://orcid.org/0000-0002-2372-9502}.
\end{bmhead}

\begin{appen}
\section{Estimation of uncertainty}\label{uncertainty_estimation}
Accurate quantification of wall-pressure fluctuations is critical for characterizing turbulent noise in this study. Uncertainties are estimated following \citet{Joseph2020JFMPlate}, combining Type A (random, statistical) and Type B (systematic, calibration-based) contributions with a coverage factor $k=2$ (95\% confidence). These are incorporated into data processing and validated against benchmarks. Uncertainty sources and calculations are as follows:

\begin{enumerate}[label=(\roman*)]
	\item Sensor measurement: The CYG1506G-P4LS12C2 transducers have a specified accuracy of $\pm 0.5$\% full scale ($\pm 2$\,kPa), yielding a Type B uncertainty of $\pm 0.01$\,kPa. This assumes a uniform distribution: standard uncertainty $= 0.01 / \sqrt{3} \approx 0.0058$\,kPa, scaled by $k=2$.
	
	\item Calibration and transfer function: Helmholtz correction (\S\ref{Wall-pressure_signal_correction}, Step 1) uses a fitted transfer function. Residuals from figure~\ref{fig:frequency_response} give an RMSE of 0.0092 over 0 -- 10\,kHz. This Type B uncertainty converts to $\pm 0.2$\,dB at 95\% confidence: $20 \log_{10}(1 + 2 \times 0.0092) \approx 0.16$\,dB, conservatively rounded.
	
	\item Noise cancellation: Wiener filter uncertainty (\S\ref{Wall-pressure_signal_correction}, Step 2) is assessed via jitter in filter order ($m=16000 \pm 1000$, 10 trials). PSD variation at 1\,kHz yields $\sigma \approx 0.5$\,dB (Type A). Standard error: $u_{\text{noise}} = 0.5 / \sqrt{10} \approx 0.16$\,dB, scaled by $k=2$.
	
	\item Sampling and spectral processing: With 20\,kHz sampling and 511 segments of 8192 points (\S\ref{Wall-pressure_signal_correction}, Step 3), averaging reduces random error to $u_{\text{avg}} = \sigma_{\text{PSD}} / \sqrt{511} < 0.1$\% (negligible). Hanning windowing introduces $\pm 1$\,dB systematic uncertainty, consistent with Joseph \textit{et al.}.
\end{enumerate}

\begin{table}
	\centering
    \def~{\hphantom{0}}
	\begin{tabular}{lcc}
		Parameter & Uncertainty & Source \\
		Static pressure $p$ & $\pm 0.05$\,Pa & Transducer accuracy (i) \\
		Wall-pressure fluctuation $p'$ & $\pm 0.01$\,kPa & Sensor (i) \\
		RMS pressure $p_{\text{rms}}$ & $\pm 3$\% & PSD integration (ii--iv) \\
		PSD $\Phi(f)$ (low frequency) & $\pm 2.0$\,dB & Jitter, noise, calibration (ii--iv) \\
		PSD $\Phi(f)$ (high frequency) & $\pm 1.0$\,dB & Calibration residuals (ii) \\
	\end{tabular}
    \caption{Summary of uncertainties (95\% confidence).}
	\label{tab:uncertainty_summary}
\end{table}

Uncertainty propagation: For RMS pressure ($p_{\text{rms}} = \sqrt{\langle p'^2 \rangle}$), relative uncertainty is $u_{p_{\text{rms}}} / p_{\text{rms}} = 0.5 \cdot u_{\Phi} / \Phi$. Jitter ($\pm 2$\,dB) corresponds to $\pm 25$\%; calibration ($\pm 0.2$\,dB) to $\pm 2.3$\%. Combined: $u_{\Phi} \approx \sqrt{0.25^2 + 0.023^2} \approx 25.1$\%, yielding $\pm 3$\% for $p_{\text{rms}}$. For PSD, $u_{\Phi} = \sqrt{u_{\text{jitter}}^2 + u_{\text{calib}}^2 + u_{\text{noise}}^2}$ gives $\pm 2.0$\,dB (low frequency) and $\pm 1.0$\,dB (high frequency), the latter reduced by precise fitting. Table~\ref{tab:uncertainty_summary} summarizes these estimates.

\end{appen} 

\bibliographystyle{jfm}

\end{document}